\documentclass[sigconf]{acmart}
\AtBeginDocument{%
  }

\copyrightyear{2026}
\acmYear{2026}
\setcopyright{cc}
\setcctype{by}
\acmConference[CHI '26]{Proceedings of the 2026 CHI Conference on Human Factors in Computing Systems}{April 13--17, 2026}{Barcelona, Spain}
\acmBooktitle{Proceedings of the 2026 CHI Conference on Human Factors in Computing Systems (CHI '26), April 13--17, 2026, Barcelona, Spain}
\acmDOI{10.1145/3772318.3790710}
\acmISBN{979-8-4007-2278-3/2026/04}

\usepackage{tabularx, colortbl}
\usepackage{multirow, makecell}

\usepackage{enumitem}

\usepackage[T1]{fontenc}
\usepackage{stfloats}
\usepackage{array}
\usepackage{listings}
\usepackage{amsmath}
\usepackage{xspace}
\newcommand{\etal}{\textit{et al.}}

\newcommand{\para}[1]{\noindent\textbf{#1}}

\newcommand{\carton}[1]{{\color{black}{#1}}}
\newcommand{\final}[1]{{\color{black}{#1}}}

\newcommand{\AI}{\texttt{AI}\xspace}
\newcommand{\noAI}{\texttt{NoAI}\xspace}
\newcommand{\AR}{\texttt{X-ray}\xspace}
\newcommand{\map}{\texttt{Minimap}\xspace}

\newcommand{\ARAI}{\texttt{X-ray}\allowbreak+\allowbreak\texttt{AI}\xspace}
\newcommand{\ARnoAI}{\texttt{X-ray}\allowbreak+\allowbreak\texttt{NoAI}\xspace}
\newcommand{\mapAI}{\texttt{Minimap}\allowbreak+\allowbreak\texttt{AI}\xspace}
\newcommand{\mapnoAI}{\texttt{Minimap}\allowbreak+\allowbreak\texttt{NoAI}\xspace}

\newcommand{\SC}{\textsc{Same+Close}\xspace}
\newcommand{\SF}{\textsc{Same+Far}\xspace}
\newcommand{\CC}{\textsc{Cross+Close}\xspace}
\newcommand{\CF}{\textsc{Cross+Far}\xspace}

\newcommand{\quot}[1]{\emph{``#1''}}

\begin{document}

%%
%% The "title" command has an optional parameter,
%% allowing the author to define a "short title" to be used in page headers.

% \title{Can AR-Embedded Visualizations Foster Appropriate Reliance on AI in Spatial decision-making?
% A Comparative Study of AR-Embedded vs. 2D Minimap}

% \title[Can AR Embedded Visualizations Foster Appropriate Reliance on AI in Spatial decision-making?]{Can AR Embedded Visualizations Foster Appropriate Reliance on AI in Spatial decision-making?
% A Comparative Study of Embedded AR vs. 2D Minimap}

\title[Can AR Embedded Visualizations Foster Appropriate Reliance on AI in Spatial Decision-Making?]{Can AR Embedded Visualizations Foster Appropriate Reliance on AI in Spatial Decision-Making? \\ A Comparative Study of AR X-Ray vs. 2D Minimap}

%%
%% The "author" command and its associated commands are used to define
%% the authors and their affiliations.
%% Of note is the shared affiliation of the first two authors, and the
%% "authornote" and "authornotemark" commands
%% used to denote shared contribution to the research.
\author{Xianhao Carton Liu}
\email{liu03008@umn.edu}
\orcid{0009-0006-3528-9651}
\affiliation{%
  \institution{University of Minnesota}
  \city{Minneapolis}
  \state{Minnesota}
  \country{USA}}

\author{Difan Jia}
\email{bobby003@umn.edu}
\orcid{0009-0008-7470-5202}
\affiliation{%
  \institution{University of Minnesota}
  \city{Minneapolis}
  \state{Minnesota}
  \country{USA}
}

\author{Tongyu Nie}
\email{nie00035@umn.edu}
\orcid{0000-0003-4186-8749}
\affiliation{%
  \institution{University of Minnesota}
  \city{Minneapolis}
  \state{Minnesota}
  \country{USA}
}

\author{Evan Suma Rosenberg}
\email{suma@umn.edu}
\orcid{0000-0002-4826-4561}
\affiliation{%
  \institution{University of Minnesota}
  \city{Minneapolis}
  \state{Minnesota}
  \country{USA}
}

\author{Victoria Interrante}
\email{interran@umn.edu}
\orcid{0000-0002-3313-6663}
\affiliation{%
  \institution{University of Minnesota}
  \city{Minneapolis}
  \state{Minnesota}
  \country{USA}
}

\author{Chen Zhu-Tian}
\email{ztchen@umn.edu}
\orcid{0000-0002-2313-0612}
\affiliation{%
  \institution{University of Minnesota}
  \city{Minneapolis}
  \state{Minnesota}
  \country{USA}
}
%%
%% By default, the full list of authors will be used in the page
%% headers. Often, this list is too long, and will overlap
%% other information printed in the page headers. This command allows
%% the author to define a more concise list
%% of authors' names for this purpose.
\renewcommand{\shortauthors}{Liu et al.}

%%
%% The abstract is a short summary of the work to be presented in the
%% article.
\begin{abstract}
Artificial Intelligence (AI) and indoor sensing increasingly support decision-making in spatial environments.
However, traditional visualization methods impose a \carton{substantial} mental workload when viewers translate this digital information into real-world spaces, leading to inappropriate reliance on AI. Embedded visualizations in Augmented Reality (AR), by integrating information into physical environments, may reduce this workload and foster more appropriate reliance on AI. 
To assess this, we conducted an empirical study (N = 32) comparing an AR embedded visualization (X-ray) and 2D Minimap in AI-assisted, time-critical spatial target selection tasks. Surprisingly, evidence shows that the embedded visualization led to greater inappropriate reliance on AI, primarily as over-reliance, 
due to factors like perceptual challenges, visual proximity illusions, and highly realistic visual representations.
Nonetheless, the embedded visualization showed benefits in spatial mapping. We conclude by discussing empirical insights, design implications, and directions for future research on human-AI collaborative decision in AR.

\end{abstract}

%%
%% The code below is generated by the tool at http://dl.acm.org/ccs.cfm.
%% Please copy and paste the code instead of the example below.
%%


%%
%% Keywords. The author(s) should pick words that accurately describe
%% the work being presented. Separate the keywords with commas.
%% A "teaser" image appears between the author and affiliation
%% information and the body of the document, and typically spans the
%% page.
\begin{teaserfigure}
  \vspace{-3mm}
  \includegraphics[width=\linewidth]{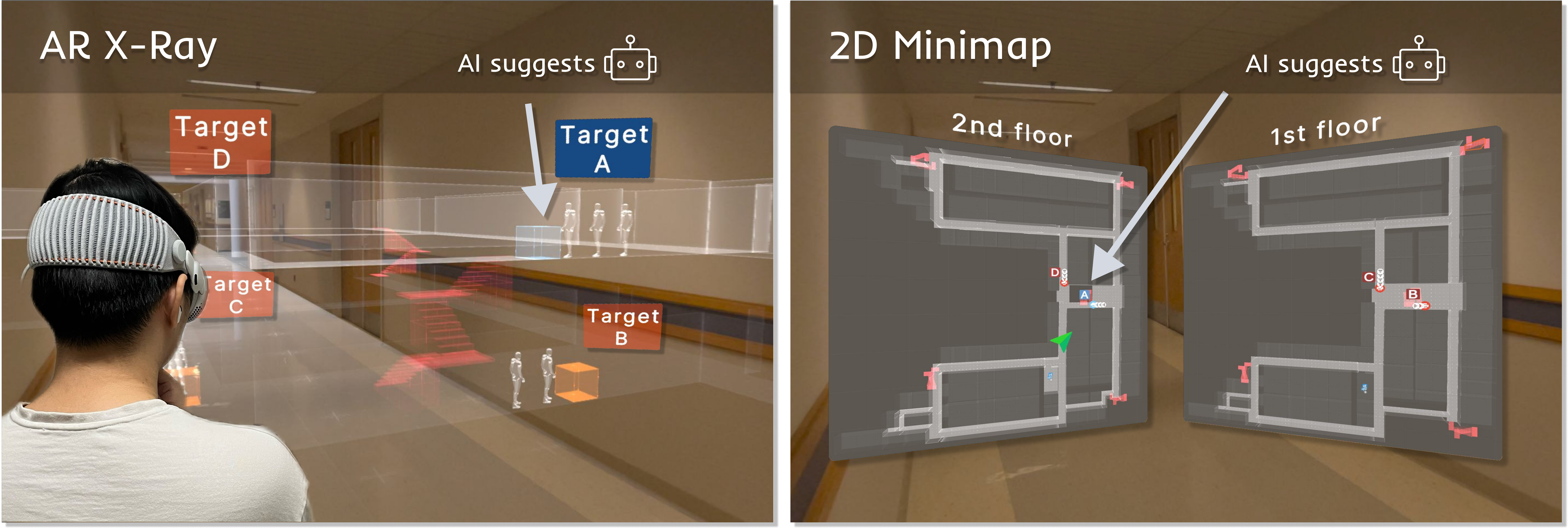}
  \vspace{-7mm}
  \caption{A participant wearing an Apple Vision Pro performs spatial decision-making tasks using two different visualizations: AR X-ray (Left) and 2D Minimap (Right). Both visualizations present AI-suggested targets, but the AR X-ray approach embeds information directly into the participant's field of view, while the 2D Minimap offers a top-down abstraction. 
  % The depiction of targets across two floors is for illustration only and does not reflect the actual experimental setup.
  }
  \Description{A participant wearing an Apple Vision Pro performs spatial decision-making tasks using two different visualizations: AR X-ray (Left) and 2D Minimap (Right). Both visualizations present AI-suggested targets, but the AR X-ray approach embeds information directly into the participant's field of view, while the 2D Minimap offers a top-down abstraction. The depiction of targets across two floors is for illustration only and does not reflect the actual experimental setup.}
  \label{fig:teaser}
\end{teaserfigure}

% \received{20 February 2007}
% \received[revised]{12 March 2009}
% \received[accepted]{5 June 2009}

%%
%% This command processes the author and affiliation and title
%% information and builds the first part of the formatted document.

\maketitle
\section{Introduction}

\carton{From high-stakes emergency evacuations~\cite{Tim2022interactiveEvacuation, Manfredi2024AMR} and crisis management~\cite{Launder2014ASI}, to everyday situations like hurrying to a classroom or navigating an unfamiliar building under time pressure,} people often have to rapidly choose among multiple spatial targets \carton{such as} exits, areas to secure, and hallways.
We refer to these decision-making tasks, which require integrating information about spatial environments with external data, as \emph{Spatial Decision-Making}. 
\carton{Making a poor choice can waste time and resources, and in the worst cases,}
endanger lives~\cite{henig1996solving}.
Recent advancements in indoor sensing and artificial intelligence (AI) offer new avenues for supporting such decisions by visualizing real-time situational data and AI suggestions in a spatial context~\cite{XU2021Seethroughwalls,Nick2024OminiActions}.

However, the way this information is presented can significantly impact user behavior and decision quality~\cite{Gaba2024ModelUnFairPeopleCareOrNot}.
Existing methods~\cite{li2023autonomousgisnextgenerationaipowered, Xu2024EvaluatingLL} commonly employ 2D maps to display data and AI suggestions, requiring users to mentally map and connect these digital cues onto physical spaces.
According to spatial cognition research~\cite{Klatzky1998AllocentricAE}, this process, known as \emph{reference frame translation}, can impose a \carton{substantial} cognitive load.
Under time pressure, users may struggle to critically assess AI outputs, 
resulting in \emph{inappropriate reliance}~\cite{Swaroop2024AccuracyTime}: either trusting flawed AI suggestions or rejecting correct ones. 
This concern is especially pressing in emergency contexts, where every second counts.

% Depending on the
% spatial proximity between visualization and referent, Willett \etal [115]
% classify visualizations into non-situated (\ie, in different locations),
% situated (\ie, in the same location), and embedded (\ie, directly on top
% or adjacent to the referent). S

Augmented Reality (AR) promises a more intuitive approach by directly visualizing data and AI suggestions onto the physical environment.
These visualizations, often referred to as \emph{embedded visualizations}\final{~\cite{Willett2017EmbeddedDataRepresentations, chen2023iball}}, can reduce cognitive load by eliminating the need for constant reference frame translation.
Intuitively, embedded visualizations might foster more deliberate and appropriate reliance on AI. 
% , more than \emph{situated visualizations} which displays data in the same location as the physical referent,
% -- often referred to as \emph{Situated Visualizations}\cite{lee2023design, Bressa2022SituatedVisualization, Shin2024TheRealityofTheSituation} or \emph{Embedded Visualizations}\cite{Willett2017EmbeddedDataRepresentations} -- 
Yet, whether embedded visualizations genuinely mitigates inappropriate reliance on AI in spatial decision-making remains unclear, and little existing work has examined how embedded visualizations in AR affects human-AI reliance in spatial tasks.
As AR and AI continue to mature and increasingly integrate~\cite{Hirzle2023WhenXRAndAIMeet}, 
this work aims to provide researchers and practitioners with initial evidence on how embedded visualizations impact human-AI reliance in spatial decision-making. 

% To this end, we present an empirical study (N=32) comparing an AR embedded visualization (\autoref{fig:teaser} Left, ``X-ray'' view) against a 2D Minimap (\autoref{fig:teaser} Right) in a time-critical, AI-assisted spatial decision task.
To this end, we present an empirical study (N=32) comparing an AR X-ray (\autoref{fig:teaser} Left) against a 2D Minimap (\autoref{fig:teaser} Right) in a time-pressured, AI-assisted spatial decision task.
Following previous work on AR visual systems for indoor navigation\final{~\cite{XU2021Seethroughwalls}}, we selected the X-ray view as our AR embedded visualization because it naturally embeds information within a large-scale indoor environment.
In the tasks, participants needed to select one of four targets distributed across a two-floor building based on multiple criteria under time pressure.
An AI with 75\% accuracy was provided to suggest the optimal target, and both visualizations were experienced using an Apple Vision Pro (AVP)~\footnote{\url{https://www.apple.com/apple-vision-pro/}}, a state-of-the-art AR headset. 
\carton{This abstract task models time-pressured spatial choices with imperfect AI support, where over-reliance can hide errors and under-reliance can waste useful guidance.}

Contrary to our expectations, participants exhibited greater inappropriate reliance on AI in the \carton{X-ray} condition, suggesting that new perceptual and cognitive challenges arise when digital cues are spatially embedded in large-scale physical environments. 
Further analysis of quantitative data (decision accuracy, response times, reliance metrics, and self-reported confidence) reveals that inappropriate reliance manifested primarily as over-reliance, \final{which means,} blindly accepting AI suggestions. 
Through qualitative interviews, we identify several key factors that contributed to the over-reliance, such as perceptual challenges in AR, visual proximity illusions, and heightened trust in ``embodied'' AI suggestions.
Nonetheless, our findings suggest \carton{the X-ray}'s strength in spatial mapping (as evidenced by quantitative data) and support for egocentric spatial imagery (as reported by participants). 
\carton{These findings matter for AR+AI design because they show that while embedding AI cues into the physical world can strengthen spatial mapping, it can also simultaneously miscalibrate reliance.}
We conclude by outlining the lessons learned, design implications, and promising future directions for better harnessing these AR strengths to foster human–AI collaboration in spatial decision-making.

% In Summary, our main contributions include:
% \begin{itemize}[leftmargin=*, noitemsep]
 
%     \item We conduct \textbf{a controlled user study with 32 participants (N=32)} comparing AR embedded visualizations and 2D minimap on a time-critical, AI-assisted spatial decision task, combining quantitative and qualitative analyses.  
%     \item We evaluate in \textbf{a large-scale real-world indoor environment} ($\approx 80 \times 80\,\text{m}$, 2 floors) using an \textbf{immersive AR platform} (\ie, Apple Vision Pro), extending beyond settings of prior work (\eg, demonstration video, mobile AR and room-scale environment). 
%     \item We uncover \textbf{empirical insights, design implications, important directions} for future research on human-AI decision collaboration in AR.
% \end{itemize}

\section{Related Work}

\subsection{Human-AI Decision-Making}
%% What is human-AI team complementary, why we need appropriate reliance 
Human-AI decision-making refers to collaborative decision processes where a human decision-maker interacts with an AI system to make a choice~\cite{2021BansalDoesTheWholeExceedItsParts}.
A common goal in these systems is to achieve \emph{complementary performance}, 
where the human-AI team performs better than either the human or the AI alone~\cite{2021BansalDoesTheWholeExceedItsParts}.
However, empirical studies have repeatedly found that such complementary performance is rarely observed ~\cite{beede2020human, carton2020feature}. 
One widely reported barrier is \emph{inappropriate reliance} — the inability of human users to correctly accept AI suggestions when they are right, or reject them when they are wrong~\cite{2021BansalDoesTheWholeExceedItsParts}.
% 2, 29, 47, 70, 83 ma2023should,schemmer2023appropriate,wang2021explanations
% in “Are You Really Sure?” Understanding the Efects of Human Self-Confidence Calibration in AI-Assisted Decision Making

Unlike \emph{trust} in AI, which reflects users' subjective feedback, \emph{reliance} objectively measures whether a participant accepts or rejects the AI's suggestion~\cite{2021BansalDoesTheWholeExceedItsParts,ma2023should}, and has therefore been widely studied.
% 2, 47, 83, 89, 90 wang2021explanations,zhang2020effect,zhao2023evaluating
% in “Are You Really Sure?” Understanding the Efects of Human Self-Confidence Calibration in AI-Assisted Decision Making
To foster \emph{appropriate reliance}, researchers have proposed providing explanations of the AI's suggestions~\cite{schemmer2023AppRelianceByExplanation}.
Unfortunately, several studies found that explanations have often failed to deliver their intended benefits, but actually increase over-reliance -- a phenomenon where users uncritically accept AI suggestions, even when they are suboptimal~\cite{2021BansalDoesTheWholeExceedItsParts}.
% Does the Whole Exceed its Parts? CHI 21
Drawn on the dual-process model of cognition~\cite{Chaiklin2012ThinkingFA, JSBTDualProcessTheory2013}, 
initial efforts attributed this over-reliance to a lack of cognitive engagement with the explanation, because people often rely on heuristics (Type 1) to judge the AI's suggestions, 
rather than engaging in deeper analysis (Type 2). 
Buçinca \etal~\cite{2021ToTrustOrToThink} introduced cognitive forcing functions that encourage users to think more critically about the AI's suggestions.
Vasconcelos \etal~\cite{vasconcelos2023explanationsreduceoverrelianceai}
% Explanations Can Reduce Overreliance on AI Systems During Decision-Making, CSCW 2023
proposed a cost-benefit framework that formalizes when users choose to engage with AI explanations.
% Chen \etal~\cite{chen2023understanding}
% % Understanding the Role of Human Intuition on Reliance in Human-AI Decision-Making with Explanations
% further highlight even when engaging with the explainations,
% users' prior beliefs and heuristics
% strongly influence whether they accept or override AI suggestions.
Recently, Guo \etal~\cite{guo2024decision} developed a theoretic framework to measure AI reliance by separating reliance behavior from cognitive limitations in signal interpretation.

In the context of visualization research, automated methods to assist humans in making decisions with data have long been practiced. While a large body of work has focused on developing visual analytics systems~\cite{Saffo2024DesignSpaceOfIA}, relatively fewer studies have investigated how visualizations affect human-AI collaboration.
Among these studies, Yang \etal~\cite{yang2020visual}
% How Do Visual Explanations Foster End Users’ Appropriate Trust in Machine Learning?, 2020
explored how visual explanations can foster appropriate trust in AI systems;
% Evaluating the Impact of Human Explanation Strategies on Human-AI Visual Decision-Making, CSCW 2023
Morrison \etal~\cite{morrison2023evaluating} investigated how different human explanation strategies might impact users' reliance on AI in visual decision-making;
Gaba \etal~\cite{Gaba2024ModelUnFairPeopleCareOrNot},
% My model is unfair, do people even care
Wang \etal~\cite{wang2022extending},
% Extending the Nested Model for User-Centric XAI: A Design Study on GNN-based Drug Repurposing, 2023
and Wall \etal~\cite{wall2024trustjunk}
% Trust Junk and Evil Knobs: Calibrating Trust in AI Visualization, 2024
each focused on how interface design influences users' perceptions of model bias and performance, highlighting the powerful role of visual presentation in shaping trust in AI systems. 
Recently, 
Ha \etal~\cite{ha2024guided}
% Guided By AI: Navigating Trust, Bias, and Data Exploration in AI-Guided Visual Analytics, 2024
found that in AI-guided visual analytics systems,
users were more inclined to accept suggestions when completing a more difficult task despite the AI's lower suggestion accuracy. 
Zhao \etal~\cite{zhao2023evaluating}
% Evaluating the Impact of Uncertainty Visualization on Model Reliance, 2024
and Reyes \etal~\cite{reyes2025trusting}
% Trusting AI: does uncertainty visualization affect decision-making?, 2025
both investigated how visualizing uncertainty in model outputs influences reliance and trust. 
Taken together, these studies underscore that visual presentation profoundly influences how users evaluate AI suggestions, rely on them, and ultimately make decisions.

Unlike all these existing works, which evaluate desktop-based human-AI collaboration, 
our study contributes an initial empirical investigation of the impact of AR visualizations on human-AI collaboration in spatial decision-making tasks. 
The differences extend beyond display technique (2D monitor vs. 3D AR) to the task environment (desktop vs. physical environment) and to cognitive processes (non-spatial vs. spatial cognition), opening new questions that merit dedicated study.

\subsection{Visualization-Aided Decision-Making}
% Decision-making has long been a central topic across multiple disciplines, including psychology~\cite{}, economics~\cite{}, management~\cite{}, and computer science~\cite{}. 
In the field of visualization, research on decision-making typically falls into two categories: understanding how humans use visualizations to make decisions\final{~\cite{padilla2018decision}} and design interactive systems ~\cite{Oral_2023_ACriticalInquiry}.

The former body of work has drawn from theories in cognitive psychology to model and explain the mechanisms by which visualizations influence decision-making. 
For example, Padilla \etal~\cite{padilla2018decision}
% Decision making with visualizations: a cognitive framework across disciplines, 2018
proposed a cognitive framework grounded in dual-process theories, illustrating how visualizations engage both intuitive (Type 1) and analytical (Type 2) reasoning processes in different contexts. 
Bancilhon \etal~\cite{bancilhon2023evaluating}
% Evaluating Visualization Decision-Making with Cognitive Models, 2023
expanded on this perspective by advocating for the integration of cognitive models in evaluating visualization effectiveness. 
A parallel line of work investigates how visualizations interact with human cognitive biases. 
Dimara \etal~\cite{dimara2018task}
% A Task-Based Taxonomy of Cognitive Biases for Information Visualization, Evanthia Dimara, 2020
introduced a task-based taxonomy of cognitive biases relevant to visualization, while Wall \etal~\cite{wall2021left}
% Left, Right, and Gender: Exploring Interaction Traces to Mitigate Human Biases, Wall2021
explored how interaction traces might mitigate such biases. 
Bearfield \etal~\cite{bearfield2024same}
% Same Data, Diverging Perspectives: The Power of Visualizations to Elicit Competing Interpretations, PacificVIS
further demonstrated that the same data visualization can elicit diverging interpretations. 
Research on uncertainty representations also plays an important role in understanding decision-making: 
Kale \etal~\cite{kale2020visual}
% Visual Reasoning Strategies for Effect Size Judgments and Decisions, VIS
showed that design choices, such as emphasizing means, can bias effect size judgments, while Fernandes \etal~\cite{fernandes2018uncertainty}
% Uncertainty Displays Using Quantile Dotplots or CDFs Improve Transit Decision-Making, 2018 CHI
found that quantile dotplots can significantly improve real-world decisions in uncertain scenarios.

The latter body of work focuses on developing interactive visualization tools to support decision-making. 
%% domain agnostic tools 
Some of these systems are domain-agnostic and designed to facilitate specific decision processes, such as inspecting rankings~\cite{wall2017podium}
% Podium: Ranking Data Using Mixed-Initiative Visual Analytics
% LineUp: Visual Analysis of Multi-Attribute Rankings
and comparing alternatives based on multiple, often conflicting attributes~\cite{pajer2016weightlifter}
% - WeightLifter: Visual Weight Space Exploration for Multi-Criteria Decision Making
%-  Solving MCDM problems: Process concepts
in Multi-Criteria Decision-Making.
Another line of work targets domain-specific decision contexts, such as medical treatment planning~\cite{kreiser2017decision}
% - Decision Graph Embedding for High-Resolution Manometry Diagnosis,
% - Visualizing and Comparing Machine Learning Predictions to Improve Human-AI Teaming on the Example of Cell Lineage
, urban policy~\cite{weng2018homefinder}
% - HomeFinder Revisited: Finding Ideal Homes with Reachability-Centric Multi-Criteria Decision Making
% - SmartAdP: Visual Analytics of Large-scale Taxi Trajectories for Selecting Billboard Locations
, and more~\cite{chang2007wirevis}.
% - Wirevis: Visualization of categorical, time-varying data from financial transactions
% - Many plans: Multidimensional ensembles for visual decision support in flood management
% - Litevis: integrated visualization for simulation-based decision support in lighting design
These systems typically incorporate domain-specific constraints to guide and contextualize the decision process.
A recent survey by Oral \etal~\cite{Oral_2023_ACriticalInquiry} offers a comprehensive analysis of decision-focused visualization tools and highlights the opportunities and challenges in supporting all stages of the decision-making process. 
% Oral \etal From Information to Choice: A Critical Inquiry Into Visualization Tools for Decision Making, 2023

While decision-making is widely recognized as a core goal of visualization, 
Dimara and Stasko~\cite{Dimara2022WhereDoDecisionMakingHide}
% A Critical Reflection on Visualization Research: Where Do Decision Making Tasks Hide? 2021
argue that it remains underrepresented in visualization research and task taxonomies. 
% They call for a more explicit integration of decision theory into the design and evaluation of visualizations.
Brumar \etal~\cite{brumar2024typologydecisionmakingtasksvisualization}
% A Typology of Decision-Making Tasks for Visualization, Camelia D. Brumar, 2024
recently proposed a typology to better describe decision-making tasks in visualization. 
Despite this growing interest,
% Nonetheless, existing work has predominantly focused on desktop settings,
% where decisions are often made away from action. 
decision support for spatial tasks remains an underexplored yet promising area~\cite{Goel2012DecisionmakingIA}. 
Our study takes a first step toward investigating spatial decision support,
aiming to provide a baseline and reference point for future research.

\subsection{Spatial Cognition and Spatial Decision-Making}
Spatial cognition encompasses the mental processes involved in perceiving, encoding, and reasoning about spatial environments~\cite{Tversky_2008_SpatialCognition}. 
A fundamental aspect of spatial cognition is the use of reference frames—mental models that help individuals locate and orient themselves in space\final{~\cite{Barbara2003StructureOfMentalSpace}}.
% Spatial Cognition, Embodied and Situated, Barbara Tversky
% Structures of mental spaces: How people think about space, Barbara Tversky
Two commonly described reference frames are \emph{egocentric}, where objects are located relative to the observer's current position and orientation \final{like} ``to my left or right'', and \emph{allocentric}, where the environment is represented independently of the observer's position, \carton{such as} map-like or object-to-object relationships~\cite{Klatzky1998AllocentricAE}.
% Allocentric and Egocentric Spatial Representations: Definitions, Distinctions, and Interconnections
People switch flexibly between these frames, but doing so can introduce \carton{substantial} cognitive load~\cite{Barbara2003StructureOfMentalSpace}, particularly when tasks require precise mental transformations, such as rotating or scaling spatial representations in mind.
% Structures of mental spaces: How people think about space, Barbara Tversky

% %% our definition of spatial decision making
Spatial decision-making requires integrating information about physical environments with external data.
To make such decisions with AI support, 
decision-makers must reconcile digital information with real-world contexts. 
With traditional 2D map-based visualizations, 
this process often involves switching between reference frames,
which is cognitively demanding and can affect interpreting or verifying AI suggestions. 
By contrast, AR can embed information directly in the physical environment, thereby eliminating the spatial mapping overhead~\cite{Billinghurst2015ARSurvey}. 
Thus, there is a sustained interest in leveraging AR systems to support spatial decision-making tasks.

\subsection{AR, Situated, and Embedded Visualizations}
Different from Virtual Reality (VR)~\cite{burdea2003virtual}, AR can directly visualize information within physical context, benefiting a broad range of applications that require computational support~\cite{Kalkofen2009ComARVis}. 
The visualization community has explored AR visualizations across multiple dimensions, including design patterns\final{~\cite{lee2023design}}, techniques
%\cite{patnaik2024vistorch, quijano2024brushing, DBLP:journals/IEEE_TVCG/YaoBVI22, assor2023handling, Zhu_Tian_2024}
\final{~\cite{Zhu_Tian_2024}}, interactions
%~\cite{fleck2022ragrug, white2009sitelens, liu2025reality}
\final{~\cite{liu2025reality, zhu2022sporthesia}}, empirical evaluations%~\cite{quijano2024brushing, merino2020toward, reitberger2007enhancing, Pietschmann2023QuantifyingTI, de2019defamiliarization}
\final{~\cite{quijano2024brushing, Pietschmann2023QuantifyingTI}}, and applications \final{~\cite{ lee2024sportify, Lin2023VIRDIM}}.
%~\cite{yang2025implementation, yu2024persival, Lin_2021, lee2024sportify}
We refer readers to the recent survey\final{~\cite{Shin2024TheRealityofTheSituation}} for more details.

More broadly, the concept of visualizing information in its physical context is known as \emph{Situated Visualization}\final{~\cite{Bressa2022SituatedVisualization}}, for which AR is one prominent realization among others \carton{like} physicalization ~\cite{Jansen2015DataPhysical}.
Depending on the spatial relationship between the visualization and its physical referent, 
Willett \etal~\cite{Willett2017EmbeddedDataRepresentations}
distinguish between \emph{situated} (co‑located in physical space) and \emph{embedded} (overlaid or integrated with the referent). 
Based on this definition, our study compares two AR visualizations: 
a \emph{situated} 2D Minimap and an \emph{embedded} AR X-ray view.
This comparison enables us to investigate whether embedding data directly into the physical referent can foster more appropriate reliance on AI by reducing the mental effort required to map data from a separate panel onto the physical environment.
For clarity in the remainder of the paper, 
\textbf{we refer to these two AR visualizations simply as \emph{Minimap} and \emph{X-ray}}, respectively.

\subsection{AR Visualizations for Spatial Decision-Making}
Our particular interest lies in leveraging AR to support indoor spatial decision-making under time pressure, a crucial area with multiple applications such as emergency evacuation~\cite{Tim2022interactiveEvacuation, Sharma2020SAARIevacuation}, first response operations~\cite{zhang2024OSTARFirstResponse}, and security management~\cite{Su2024RASSAR}.
Despite its importance, research specifically addressing time-constraint spatial decision-making in AR remains limited.
% Perhaps the closest
To the best of our knowledge, the most relevant work involves AR-based evacuation assistance,
which provides users with dynamic, context-sensitive navigation cues to safely guide them toward exits during emergencies~\cite{zhang2024OSTARFirstResponse}. 
For instance, Wächter \etal~\cite{Tim2022interactiveEvacuation} investigated AR-guided evacuation in buildings, showing that adapting AR visualizations to environmental hazards can improve evacuation efficiency and safety. 
Similarly, Sharma \etal~\cite{Sharma2020SAARIevacuation} developed AR modules to enhance situational awareness and cognitive mapping during evacuations in buildings.

Beyond emergency scenarios, a broadly related line of work explores AR-based indoor navigation, where users typically follow predefined routes rather than making spatial decisions.
Most AR navigation techniques fall into two categories: 
annotation-based methods \carton{like} arrows\carton{,} and embedded visualizations. 
Annotations are arguably the most common AR navigation technique, as seen in applications like Google Maps\footnote{\url{https://maps.google.com}}. 
Numerous studies~\cite{Lee2022NavigationInsturctions, dong2021difference} have compared annotation-based AR navigation to traditional 2D Minimap across various contexts. 
Nonetheless, AR embedded visualizations offer benefits like a global view but have received less attention, partly due to technical challenges requiring the development of digital twins of real-world environments. 
Representative work in this direction includes Xu \etal~\cite{XUImroveIndoorWayfinding, XUARteambasedsearch} who conducted comparative studies of AR X-ray views (a type of AR embedded visualizations) for indoor wayfinding, demonstrating that egocentric perspectives significantly enhanced navigation efficiency, reduced cognitive load, and improved spatial awareness.

Drawing inspiration from these works, we adopt AR X-ray views to present spatial information within indoor environments.
Unlike previous studies, our research specifically investigates
how this AR embedded visualization can support users in making spatial
decisions with AI assistance – a topic that has not yet been explored
in depth. Our study thus aims to provide an initial reference point for future research into AI-aided decision-making in AR contexts.

\section{Study Rationale}
This section defines the scope and abstraction of the spatial decision-making task, 
and summarizes the rationale behind our choices of visualizations, decision supports, 
and study hypotheses.

% \subsection{Task Motivation and Abstraction} 
\subsection{Task Abstraction and Definition} 
\label{sec:task}

%% what is spatial decision making
Spatial decision-making has long been studied under the broader umbrella of \emph{wayfinding} in spatial cognition~\cite{
Emo2012WayfindingAS, Wiener2009TaxonomyOH}.
% Taxonomy of Human Wayfinding Tasks: A Knowledge-Based Approach,
% Wayfinding and spatial configuration: evidence from street corners
% What determines our navigational abilities?
% Montello, D. (2001). Spatial Cognition. In  N. J. Smelser and P. B. Baltes (Eds.), International Encyclopedia of  the Social & Behavioral Sciences
% Wolbers2010WhatDO, Montello2001-MONSC, Meilinger2008WorkingMeminWayfinding
Foundational work by Siegel and White~\cite{Siegel1975TheDO}
% The development of spatial representations of large-scale environments 1975
conceptualizes wayfinding as 
a decision process involving three forms of spatial knowledge: landmark, route, and survey. 
Motivated by real-world applications, our study particularly focuses on \textbf{target selection} \final{(landmark)}.
% In high-stakes scenarios, 
In tasks such as \carton{exploring a building, security surveillance, and} emergency response, spatial decision-making often involves identifying or selecting a target before proceeding with an action \carton{like} navigation or intervention. 
Although AR systems have frequently been investigated for supporting the action phase \carton{such as} route guidance~\cite{XUARteambasedsearch}, their potential to enhance the preceding decision-making step remains underexplored.

To address this gap, we abstract real-world scenarios into a domain-agnostic spatial target selection task, defined by the following core attributes derived from the literature:

\begin{itemize}[leftmargin=*, noitemsep]
    \item \textbf{Space - Large-scale Indoor Environment}: 
    These tasks often take place in physical spaces larger than the immediate space around a person \carton{such as} buildings or open areas~\cite{Giuliano2013IndoorLS, Guerrieri2006RFIDassistedIL}.
% Indoor Localization System for First Responders in Emergency Scenario
% RFID-ASSISTED INDOOR LOCALIZATION AND COMMUNICATION FOR FIRST RESPONDERS*
    Tversky~\cite{Barbara2003StructureOfMentalSpace} characterize such spaces as \emph{space of navigation}.
    In this study, we focus on indoor building environments, 
    given their prevalence across numerous application domains.

    \item \textbf{Time - Sensitive}: 
    % The users need 
    % prompt decision-making to mitigate risks and ensure task effectiveness
    Many high-stakes spatial tasks impose stringent time constraints, requiring users to make optimal decision as soon as possible~\cite{nydegger2011post}.
% Posttraumatic stress reactions in volunteer firefighters
% Post-Traumatic Stress Disorder And Coping Among Career Professional Firefighters
    In this study, we focus on time-pressure scenarios that particularly benefit from decision support systems.

    \item \textbf{Data - Static and Dynamic}: 
    Appropriate target selection relies on both static data \carton{like} building level layouts, and dynamic updates  from sensors \final{like} walking people, occupant counts, and hazard alerts to evaluate each target~\cite{Adib2013SeeThroughWallsWithWifi}.

    \item \textbf{Decision - Multi-Criteria}: 
    These tasks frequently require decisions among multiple targets or locations distributed over rooms, floors, or areas~\cite{Launder2014ASI}. 
% A study identifying factors influencing decision making in dynamic emergencies like urban fire and rescue settings
    Success often involves balancing or prioritizing several factors, 
    such as response time, travel distance, and urgency~\cite{penney2022threat}.
% Threat assessment, sense making, and critical decision-making in police, military, ambulance, and fire services
       
\end{itemize}
% Table~\ref{} summarizes these core elements, along with their real-world analogs in relevant application domains.
By focusing on these domain-agnostic factors, we can investigate how AR-based decision support influences spatial decision-making without limiting our findings to one specific use case.

\subsection{Visualization Methods: Minimap vs. X-Ray}
Given that spatial decision-making inherently involves interpreting spatial relationships, 
an effective decision support tool must present relevant data within its spatial context. 
Thus, we select map-based visualizations as the foundational interface for our study. 
We are particularly interested in how \textbf{embedded visualizations}
influence users' decision-making processes. 
Therefore, we choose two representative visualization methods: 
a Minimap and a X-ray.
Below, we discuss the rationale behind choosing these two methods 
and briefly highlight alternative visualization approaches.

\para{Minimap.} 
2D Minimap is perhaps the most widely adopted map-based visualization on traditional flat-screens. 
It presents the spatial environment as a simplified, abstracted, top-down representation, upon which additional data can be annotated, \carton{such as} Google Maps.
For indoor environments spanning multiple floors, 2D Minimaps often employ separate layers for each floor, indicating spatial relationships through key landmarks such as stairways or elevators.
The use of 2D Minimap is well-established and requires minimal training, making them a suitable baseline for comparison.
In our study, we implement a Minimap (\autoref{fig:teaser} Right) that shows the spatial distribution of multiple targets and their associated data using glyphs.

\para{X-Ray.}
Given our task focuses on large-scale indoor environments, 
we select AR X-ray (\autoref{fig:teaser} Left) as our embedded visualizations, which overlays a 1:1 scale 3D map on the real-world.
It allows users see the interior layouts or objects behind walls, as if they had ``X-ray'' vision~\cite{Livingston2013PursuitO}, which has been widely used in applications like navigation~\cite{XUImroveIndoorWayfinding} and construction~\cite{Soria2018AugmentedAV}.
For our study, 
we follow previous similar research\final{~\cite{XUImroveIndoorWayfinding}} and portray the building’s indoor layout, along with multiple targets (represented as cubes), 
and their associated data \final{(the number of people in line)} using virtual avatars.

\para{Alternatives.} World-in-Miniature (WiM)~\cite{Stoakley1995WIM, Danluk2021WiMs} provides a three-dimensional representation of the environment and can be implemented on flat-screen or AR displays. However, WiM typically relies heavily on interactions such as panning, tilting, zooming, or rotation to fully leverage their advantages. Because our current investigation prioritizes evaluating the impact of the visual representations themselves (without the influence of interactive complexity), we intentionally avoided such interactions in our study. Thus, we reserve exploration of interactive WiM and other advanced visualization methods for future research.

\subsection{Visual Decision Supports}
A recent survey by Oral \etal~\cite{Oral_2023_ACriticalInquiry}
highlights a broad spectrum of visual decision support tools, 
ranging from basic information visualizations to advanced AI-based suggestions~\cite{Lawless2024IWantItThatWay}. 
For this study, we specifically choose to implement two fundamental yet representative forms of decision support across both the Minimap and X-ray:
\begin{itemize}[leftmargin=*, noitemsep]
    \item \textbf{Target Visualizations} 
     that present candidate targets and associated real-time data \carton{like} occupancy counts on the map, without path suggestions or explanatory text.
      
    \item \textbf{AI-Suggestions} 
    that highlight a single optimal target based on predefined criteria, with no accompanying explanation. Similar to all AI systems, suggestions may not always be accurate.
    We detail our adjustment of the performance of the AI based on prior works~\cite{Swaroop2024AccuracyTime} and pilot studies in Sec.~\ref{sec:simulated_AI}.
\end{itemize}

\noindent
Following practices established in prior research~\cite{vasconcelos2023explanationsreduceoverrelianceai}, 
we deliberately constrain our decision support to these fundamental features for several reasons:
First, target and real-time data visualization represent core functionalities broadly used in existing decision-support tools. 
Second, simple AI suggestions enable direct investigation of user reliance on AI assistance under different visualizations, without the additional cognitive complexity and potential confounding effects introduced by detailed explanations~\cite{Jiang2018ToTrustOrNotToTrust}.

While more advanced support, such as comparative visualizations, AI-generated explanations, and uncertainty displays, may offer added benefits, we intentionally focus on these fundamental forms to isolate the role of embedded visualizations in spatial decision-making and AI reliance. 
This controlled setup provides a baseline for future studies exploring richer decision supports.

\section{Study Design}
We conducted a 2 $\times$ 2 within-subjects study to investigate whether X-ray can foster more appropriate reliance on AI in spatial decision-making. 
The two independent variables were the type of visualizations (\AR vs. \map), and the availability of AI assistance (\noAI vs. \AI). The study design is detailed below, with fixed values based on a 13-participant pilot study.
\carton{The study was approved by university ethics review board (IRB).}

\subsection{Task Description}
We designed a realistic indoor building scenario,
where participants acted as busy employees. 
The goal was to simulate a time-pressured spatial decision-making scenario that closely resembles everyday scenarios to the participants.
In each trial, participants were tasked with \textbf{selecting one coffee machine from four available options within a two-floor building within 20 seconds}. 
The objective was to choose the machine that would allow them to obtain a cup of coffee in the shortest possible time. 
Optimal decision-making in this task depended on two key factors: 
\begin{itemize}[leftmargin=*, noitemsep]
    \item \textbf{Walking Distance} — the length of the walking path from the participant's current location towards the spatial location of each coffee machine.

    \item \textbf{Queue Length} — the number of people currently waiting in line at each machine, which thus decides waiting time.
\end{itemize}
Participants accessed information about both the machine locations and queue lengths via one of two visualizations, and in half of the trials, they also received AI suggestions.

\subsection{Spatial Arrangement}
The task was set in a real two-floor indoor building spanning roughly 80m $\times$ 80m.
In each trial, the participant and the four coffee machine targets were positioned based on 
a predefined \emph{spatial arrangement}.

\begin{figure}[htb]
    \centering
    \includegraphics[width=\linewidth, keepaspectratio]{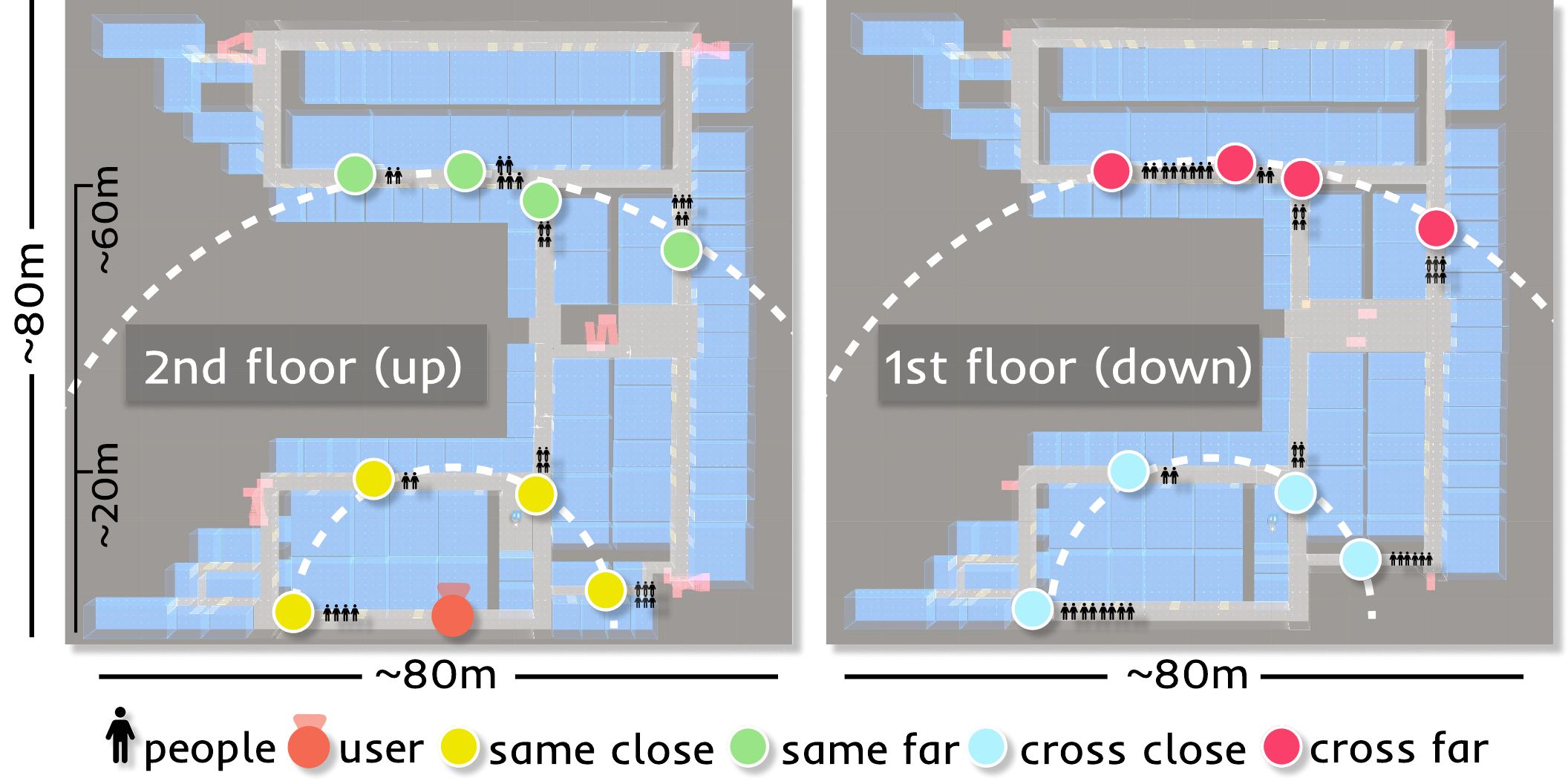}
  \caption{An example spatial arrangement \carton{of the coffee machine targets locations} for a single participant location, illustrating four distinct difficulty levels: \textsc{same}+\textsc{close}, \textsc{same}+\textsc{far}, \textsc{cross}+\textsc{close}, and \textsc{cross}+\textsc{far}. \carton{For each participant location, there are 4 trials, one difficulty level per trial.} The participant's location is marked in red, and colored dots indicate target locations in each difficulty level. \carton{The white dashed lines indicate that the coffee machine targets are placed at equal line-of-sight distances from the participant.}}
    \label{fig:MapAndTargetDesign}
    \Description{An example spatial arrangement \carton{of the coffee machine targets locations} for a single participant location, illustrating four distinct difficulty levels: \textsc{same}+\textsc{close}, \textsc{same}+\textsc{far}, \textsc{cross}+\textsc{close}, and \textsc{cross}+\textsc{far}. \carton{For each participant location, there are 4 trials, one difficulty level per trial.} The participant's location is marked in red, and colored dots indicate target locations in each difficulty level. \carton{The white dashed lines indicate that the coffee machine targets are placed at equal line-of-sight distances from the participant.}}
\end{figure}

\para{Target and Participant Locations.}
To simulate diverse spatial scenarios and mitigate learning effects, 
we varied the distribution of the four coffee machines across trials based on two spatial factors identified during the pilot testing:
\begin{itemize}[leftmargin=*, noitemsep]
    \item \textbf{Floor Relationship -- \textsc{same} or \textsc{cross}:} 
    All four targets were located on the same floor, which was either the same as the participant's floor (\textsc{same}) or a different one (\textsc{cross}).
    We anticipated that this factor would \carton{substantially} impact the difficulty of estimating walking distances, particularly in the \map condition.

    \item \textbf{Proximity -- \textsc{close} or \textsc{far}:} 
    Targets were categorized as either \textsc{close} or \textsc{far} in direct proximity (instead of walking distance) to the participant's location. 
    We anticipated that this factor would have a greater impact on task difficulty in the \AR condition, as distant targets would subtend a smaller retinal angle than closer targets and thus be harder to perceive.
\end{itemize}

The combination of these two factors resulted in four difficulty levels, 
ranging from \SC (easiest) to \CF (most difficult). 
Within each trial, to ensure that the appearance of each target was perceptually similar to the user, we used the same difficulty level setting to define each target's location and placed them all at equal line-of-sight distances from the participant's location. The participant's location was then positioned at the center of a circle passing through all four targets.
~\autoref{fig:MapAndTargetDesign} illustrates an example of the four difficulty levels.

\begin{figure}[htb]
    \centering
    \includegraphics[width=\linewidth, keepaspectratio]{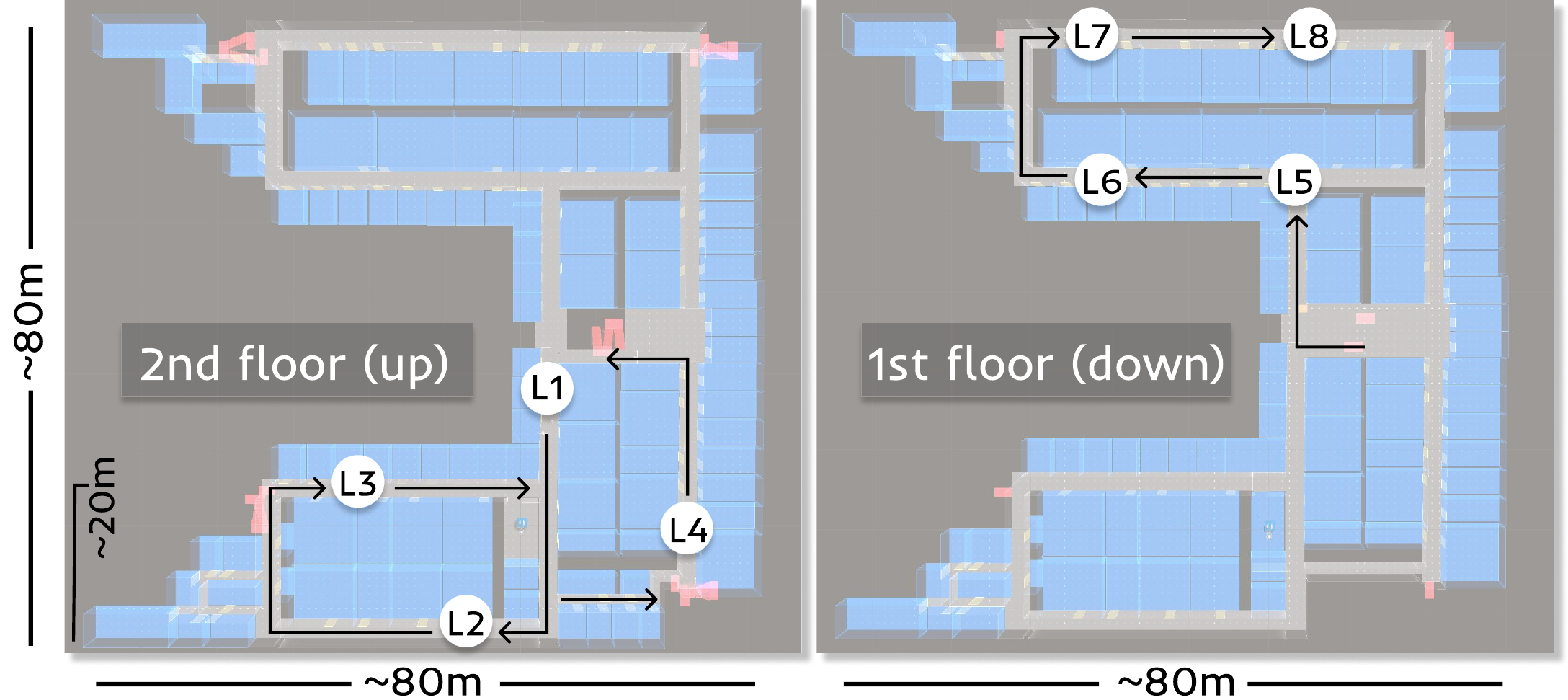}
    \caption{The 8 participant locations in the study, 
    distributed across two floors of the building. 
    Participants were instructed to complete 4 trials at each location, following the order indicated by the numbered labels \carton{and route}. The letter ``L'' is an abbreviation for location.}
    \label{fig:routeOrder}
    \Description{The 8 participant locations in the study, 
    distributed across two floors of the building. 
    Participants were instructed to complete 4 trials at each location, following the order indicated by the numbered labels \carton{and route}. The letter ``L'' is an abbreviation for location.}
\end{figure}

\para{Grouping Arrangements by Participant Location.}
Ideally, both participant and target locations would vary each trial.
However, moving participants between locations would introduce \carton{substantial} overhead and fatigue.
To address this, we grouped the target arrangements by participant location.
Specifically, we selected eight participant locations evenly distributed in the building (\autoref{fig:routeOrder}), and for each location, we designed four spatial arrangements of targets, 
each representing one difficulty level. This resulted in $4 \times 8 = 32$ unique spatial arrangements. An example of one group is shown in ~\autoref{fig:MapAndTargetDesign}.

\para{Target Optimality.}
Target optimality was determined by two metrics: 
walking distance from the participant and queue length.
For simplification,
we estimate the total time to obtain coffee from a target using this formula:
\[
T_{\text{coffee}}(d, n) = \frac{d}{\mathrm{SPEED}} + n \times \mathrm{SERVICE\_TIME}
\]
where $d$ denotes the walking distance to the target, and $n$ means the number of people in line. 
$\mathrm{SPEED}$ and $\mathrm{SERVICE\_TIME}$ are constants
denote the average walking speed and the average waiting time per person in line, respectively.
Based on our pilot studies, we empirically set $\mathrm{SPEED}$ at 1m/s and $\mathrm{SERVICE\_TIME}$ as 15s.
While individuals naturally differ in walking speed, participants in our study did not need to physically walk; $\mathrm{SPEED}$ was used for estimation and is thus not expected to introduce \carton{substantial} bias.
These parameters were explicitly communicated to participants in the study.

Since the values of $d$ were fixed by the spatial arrangement of the targets,
we manipulated the queue lengths $n$ to differentiate target optimality.
Following prior research~\cite{Swaroop2024AccuracyTime}, each trial was designed to include intentionally varied options: \textbf{one optimal, two suboptimal, and one worst choice}, differing by at least 10 seconds in $T_\text{coffee}$. 
This setup allowed us to evaluate potential participant over-reliance on AI suggestions. Detailed calculations are provided in the supplementary materials.

\subsection{Implementation and Apparatus}
Following similar work~\cite{2021ToTrustOrToThink},
a simulated AI was used to provide decision suggestions in applicable trials.
Details of the simulated AI setup are provided in Sec.~\ref{sec:simulated_AI}.
We developed the minimap based on high-precision CAD models provided by the building manager. 
The AR visualizations was developed using ARKit~\footnote{\url{https://developer.apple.com/augmented-reality/arkit/}} and AR Mesh Manager ~\footnote{\url{https://docs.unity3d.com/Packages/com.unity.xr.arkit@6.2/manual/arkit-meshing.html}} to scan the building and construct a 3D digital twin in Unity 6000.0.29f1, 
with ARAnchor Manager ~\footnote{\url{https://docs.unity3d.com/Packages/com.unity.xr.arfoundation@6.2/manual/features/anchors/aranchormanager.html}} ensuring proper registration of the digital model with the physical environment. 
The system is available at our public repository \url{https://github.com/demoPlz/ARAIReliance}. All trials were completed using an Apple Vision Pro to eliminate device-related differences and provided a high-fidelity immersive experience.

\subsection{Participants}
We recruited 32 participants (16 male, 16 female; ages 18–28) from our university community via email, word of mouth, and posted flyers. 
% We balanced the number of gender to avoid known spatial difference in gender~\cite{}.
For base compensation, participants received a \$15 Amazon gift card. To incentivize accurate performance, participants earned an additional \$0.20 for each trial in which they selected the optimal target, with the bonus capped at \$20.

To use Apple Vision Pro, participants were required to have normal vision, use contact lenses, or use the prescription inserts provided by us. To minimize potential bias arising from participants' familiarity with the experimental setting, we exclusively recruited individuals whose prior exposure to the building used in the study was limited. Specifically, 19 participants (9 male, 10 female) had never visited the building, 12 participants (6 male, 6 female) had been in the building between one and five times, and 1 participant (male) had visited the building between five and ten times.

\begin{figure*}[htb]
\centering
\includegraphics[width=\linewidth]{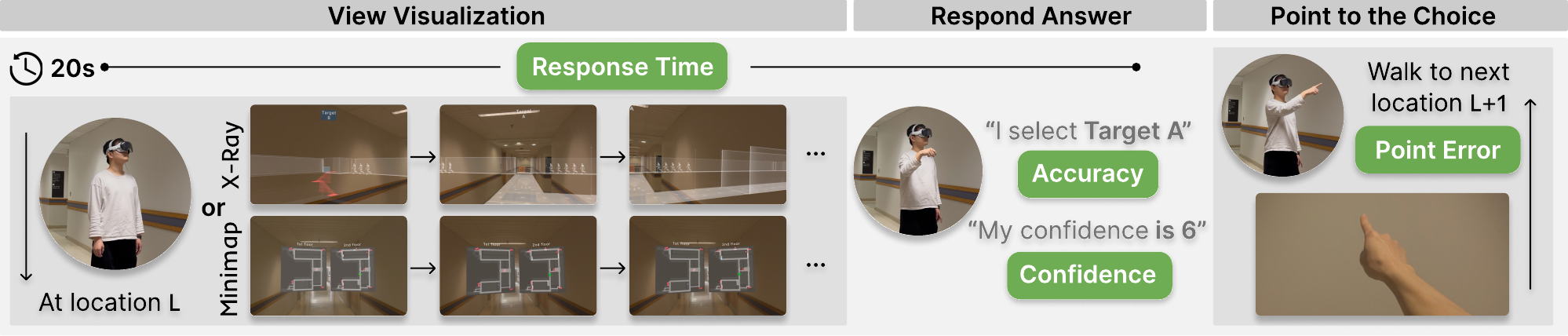}
\caption{\carton{A stimulus workflow from an actual trial. Each trial began with a 20-second countdown. Participants viewed either the \AR or \map visualization while searching for the optimal target. After identifying their choice, participants verbally announced the selected target and rated their confidence on a 1–7 scale. Finally, they physically pointed to the perceived location of the chosen target. For illustration, the figure shows only one trial at location L, but participants completed four trials at each location.}}
\Description{A stimulus workflow from an actual trial. Each trial began with a 20-second countdown. Participants viewed either the \AR or \map visualization while searching for the optimal target. After identifying their choice, participants verbally announced the selected target and rated their confidence on a 1–7 scale. Finally, they physically pointed to the perceived location of the chosen target. For illustration, the figure shows only one trial at location L, but participants completed four trials at each location.}
\label{fig:trialworkflow}
\end{figure*}

\subsection{Procedures}
With a within-subjects design, each participant completed trials under four conditions: \textbf{\mapnoAI}, \textbf{\mapAI}, \textbf{\ARnoAI}, and \textbf{\ARAI}.
The order of conditions was counterbalanced across participants using a Latin square design to minimize order effects and learning biases.
Our primary focus was on participants' behavior in the \AI conditions, 
while the \noAI conditions served as baselines to better understand user decision-making patterns.

We used the following full-factorial within-subject study design with Latin square-randomized order of the conditions
to minimize learning and fatigue effects:
% \[
% \def\arraystretch{1.2}
% \begin{array}{c c l}
%        & 32 & \text{Participants} \\
% \times & 4  & \text{Conditions: \small{\{\;\mapnoAI, \mapAI, \ARnoAI, \ARAI\}}}  \\
% \times & 2  & \text{Participant Location} \\
% \times & 4  & \text{Spatial Arrangements: \small{\{\;\SC, \SF, \CC, \CF\}}} \\
% \cline{1-3}
%        & 1024 & \text{total trials (32 per participant)}  \\
% \end{array}
% \]

\vspace{0.5\baselineskip}
\setlength{\tabcolsep}{4pt}
\renewcommand{\arraystretch}{1.2}
\small
\begin{tabular}{@{}c c p{0.8\columnwidth}@{}}
       & 32  & Participants \\
$\times$ & 4   & Conditions: {\small\{\;\mapnoAI, \mapAI, \ARnoAI, \ARAI\}} \\
$\times$ & 2   & Participant Location \\
$\times$ & 4   & Spatial Arrangements: {\small\{\;\SC, \SF, \CC, \CF\}} \\
\hline
       & 1024 & total trials (32 per participant) \\
\end{tabular}
\vspace{0.5\baselineskip}

\begin{figure}[htb]
    \centering
    \includegraphics[width=\linewidth, keepaspectratio]{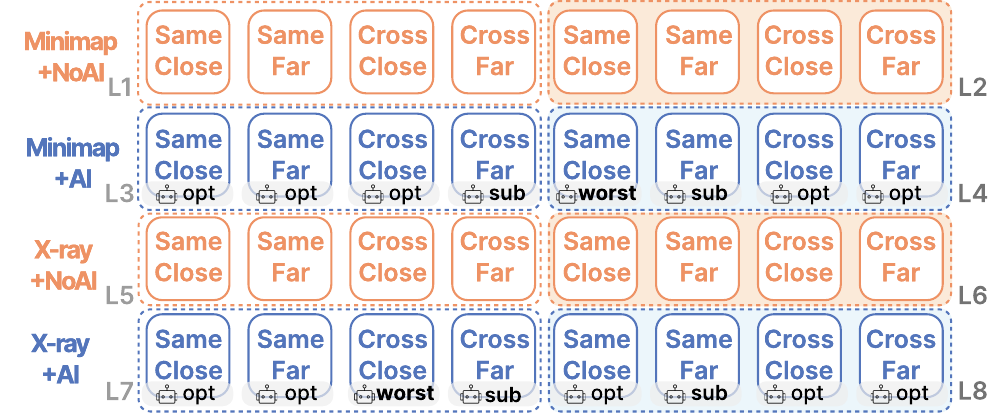}
    \caption{
    One example of what a participant experienced during the experiment. Between participants, the order of the 4 conditions (\mapnoAI, \mapAI, \ARnoAI, \ARAI) was varied and counterbalanced using a Latin square design. Factors that remained fixed across all participants included the sequence of standing locations (L1-L8) and the order of the four spatial arrangements (\SC, \SF, \CC, \CF) at each standing location. Within each AI condition, the presentation order of the AI suggestion optimality levels (opt, sub, worst) was randomized, while ensuring a balanced distribution across participants. For illustration, only two examples of AI suggestion presentation orders are shown.
    }
    \label{fig:studyMatrix}
    \Description{One example of what a participant experienced during the experiment. Between participants, the order of the conditions (rows) was varied and counterbalanced using a Latin square design across 4 conditions (\mapAI, \mapnoAI, \ARAI, \ARnoAI). Within each AI condition, the presentation order of the AI suggestion optimality levels (opt, sub, worst) was randomized. Factors that remained fixed across all participants included the sequence of standing locations (L1-L8) and the order of the four spatial arrangements (\SC, \SF, \CC, \CF) at each standing location.}
\end{figure}

Participants were split evenly between the 4 Latin square-randomized condition orders.
The orders of the spatial arrangements and participant locations were kept consistent across all participants to control for any confounding effects related to environmental layout (see \autoref{fig:studyMatrix}).

\para{Procedure Overview.}
Prior to the experiment, participants were required to take an elevator ride from the ground floor to the experiment floor (2nd floor) and wait at the elevator door for the research team. 
This procedure was designed to minimize any prior exploration or accumulation of spatial knowledge about the experimental environment. The study lasted approximately one hour in total:

\begin{itemize}[leftmargin=*, noitemsep]
    \item \emph{\underline{Introduction (10mins)}}:
    Participants first provided informed consent and were explicitly informed about the performance-based incentives. Next, participants received a detailed tutorial that explained the AVP's calibration, its usage, and task procedures.

    \item \emph{\underline{Training Tasks (10mins)}}:
    Participants completed four training trials, one per condition, to familiarize themselves with the task and visualizations. The participant locations and target locations were different in the training trials than in the experimental trials (see supplementary materials). In these trials, after participants made their selections, the optimal target was explicitly highlighted to illustrate optimal decision-making. Participants were encouraged to ask questions and were informed they could withdraw at any point.

    \item \emph{\underline{Actual Tasks (30mins)}}:
    Participants completed 32 trials, grouped into four blocks, with each block distributed over 2 consecutive locations (see the example in \autoref{fig:studyMatrix}).
    Within each block, one condition was tested, \final{namely}, eight trials per condition.  
    Within each location, they remained physically stationary and completed four trials (one trial per difficulty level).
    Breaks were allowed during the study.

    \item \emph{\underline{Post-Study Interview (10mins)}}:
    Participants were interviewed with a list of open-ended questions (attached in supplementary materials). 
\end{itemize}

\para{Detailed Trial Procedure.}
\carton{As illustrated in \autoref{fig:trialworkflow}}, each trial proceeded through the following steps:
The experimenter escorted the participant to the designated starting point and helped them put on the headset.
When ready, the experimenter used a controller to initiate the trial, 
activating the visualization and a 20-second countdown timer.
Participants verbally announced their selection, \carton{for example,} \quot{Target A}, as soon as they identified the optimal target. 
Then, the experimenter stopped the timer, automatically clearing the visualization.
Participants verbally rated their confidence in their choice using a 1 (Strongly unconfident) to 7 (Strongly confident) scale.
Lastly, participants physically pointed toward the actual spatial location of their chosen target and looked at their pointing direction. This process was repeated for each trial. We recorded the whole session through audio and first-person view video with participants' informed consent.

\subsection{Measures}
\label{sec:measures}

\begin{figure}[htb]
    \centering
    \includegraphics[width=\linewidth, keepaspectratio]{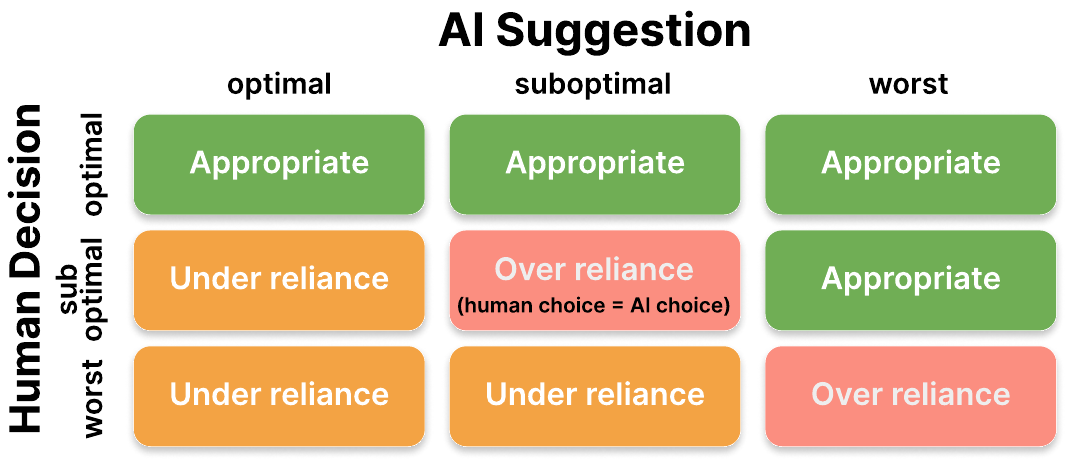}
    \caption{Classification of reliance as appropriate, over-, and under-reliance based on the alignment between AI suggestions and human decisions. Appropriate reliance on AI was defined as participants accepting optimal suggestions or overriding suboptimal and worst suggestions. Inappropriate reliance included: (1) over-reliance: accepting suboptimal or worst suggestions; and (2) under-reliance: rejecting AI suggestions in favor of a worse option.}
    \Description{Classification of reliance as appropriate, over-, and under-reliance based on the alignment between AI suggestions and human decisions. Appropriate reliance on AI was defined as participants accepting optimal suggestions or overriding suboptimal and worst suggestions. Inappropriate reliance included: (1) over-reliance: accepting suboptimal or worst suggestions; and (2) under-reliance: rejecting AI suggestions in favor of a worse option.}
    \label{fig:reliance}
\end{figure}

\noindent
For every trial, we documented and calculated the following measures based on prior research~\cite{Swaroop2024AccuracyTime, Salimzadeh2024dealWithUncertainty}:
\begin{itemize}[leftmargin=*, noitemsep]
    \item \textbf{Decision Accuracy}: 
    We defined decision accuracy as the \emph{average point score} across trials in each condition, 
    ranging from 0 to 1.
    Following previous research~\cite{Swaroop2024AccuracyTime}, participants earned points according to their choice quality:  
    optimal choice (1 point), suboptimal choices (0.5 points), and the worst choice (0 points). 
    Each trial contained exactly one optimal target, two suboptimal targets, and one worst target.

    \item  \textbf{AI Reliance}: 
    Following previous studies~\cite{yang2020visual}, we categorized participant reliance on AI as either appropriate or inappropriate (\autoref{fig:reliance}). 
    We classified reliance as appropriate when participants either followed correct AI suggestions or overrode incorrect ones with better choices.
    Inappropriate reliance included: 
    (1) over-reliance: accepting suboptimal or worst suggestions; and 
    (2) under-reliance: rejecting correct suggestions in favor of a worse option.

    \item \textbf{Response Time}: 
    We measured the duration from the start of each trial to the moment the participant verbally announced their choice.

    \item \textbf{Confidence}: 
    Participants rated their confidence in their decision on a 7-point scale from 1 (Strongly unconfident) to 7 (Strongly confident).

    \item \textbf{Point Error}: 
    Similar to prior research~\cite{hu2021comparative, warden2022visual}, 
 we assessed whether participants correctly pointed to the spatial location of their selected target based on first-person video recordings.
Two co-authors independently coded each response as 0 (correct) or 1 (incorrect) to ensure inter-rater reliability. \carton{Inter-rater reliability was evaluated via Cohen’s Kappa ($ \kappa = 0.898 $), suggesting almost perfect agreement according to Landis and Koch’s interpretation~\cite{landis1977measurement}.}
 Discrepancies were discussed and resolved through consensus. This measure indicates if participants actually knew the spatial location of their selected target.

    % Similar to prior research ~\cite{hu2021comparative, warden2022visual}, we measured this metric using participants' recorded first-person view when pointing. A response was coded as correct (1) if the real-world location of the chosen target was within the participant's field of view, and incorrect (0) otherwise. In the \map condition, the real world location may be occluded, we measured by comparing with the AR embedded visualizations of the same target setting, which provide the ground truth information of the targets' real world locations. This measure indicates whether participants correctly mapped their selected target to its physical location. 
    % More pointing errors suggest that the visualization falls short in supporting spatial mapping from the digital to the physical environment.
    
\end{itemize}

\subsection{Using Simulated AI}
\label{sec:simulated_AI}
Similar to prior works~\cite{Swaroop2024AccuracyTime, 2021ToTrustOrToThink}, we used a simulated AI with an average accuracy of 0.75 to ensure controlled experimentation. 
In each AI condition, 
the AI suggested the optimal target in five trials, a suboptimal target in two trials, and the worst target in one trial (\autoref{fig:studyMatrix}). 
This setup yields an average accuracy of 0.75 = $\frac{5 \times 1 + 2 \times 0.5 + 1 \times 0}{8}$.
The order of these suggestions was randomized and counterbalanced across participants, ensuring that optimal, suboptimal, and worst suggestions were evenly distributed across all trials and participants.

% Based on our measurement design,
% in every condition (8 trails),
% we set the AI suggested the optimal target in five trials, 
% a suboptimal target in two trials, and the worst target in one trial,
% which on average yeilds a 0.75 = 6 (\ie, 5 $\times$ 1 + 2 $\times$ 0.5 + 1 $\times$ 0) / 8 accuracy.
% The order of these suggestions was randomized and counterbalanced across participants,
% which means the AI's optimal, suboptimal, and worse suggesions are even distration in every trails across all participants.

\para{Mitigating Effects of Trust in AI.}
Following previous work~\cite{2021BansalDoesTheWholeExceedItsParts, 2021ToTrustOrToThink},
we took steps to neutralize trust effects so that observed differences in reliance could be attributed primarily to 
the visualizations.
Specifically, we avoided anthropomorphizing the AI~\cite{Kevin2015TrustInAuto}, consistently referring to it as ``the AI''. 
We also described AI outputs as ``suggestions'' to counteract perfect automation schemas~\cite{Mary2002PerceivedUtility} 
and emphasize that these suggestions could be incorrect. 
Furthermore, we do not provide decision feedback for each trial~\cite{Dietvorst2014AlgorithmAP}, or any information about the AI's accuracy~\cite{Yin2019EffectofAccOnTrust}.
% Further details regarding the AI's performance, including its counterbalance and accuracy, 
% are provided in Sec.~\ref{sec:measures} Measures.
% As a result, our simulated AI's accuracy (introduced in Sec.~\ref{}) achieved an accuracy of 0.75.

\subsection{Hypothesis}

According to dual-process theory, human decision-making can be categorized into two distinct cognitive systems: rapid, heuristic-driven \emph{Type 1} thinking, and slower, analytical \emph{Type 2} thinking~\cite{Chaiklin2012ThinkingFA}. A widely accepted distinction between these systems is that Type 2 thinking demands significantly more cognitive resources than Type 1~\cite{Chaiklin2012ThinkingFA}. 

Under conditions of high time pressure, 
users often lack sufficient cognitive bandwidth to analytically evaluate AI-generated suggestions, potentially resulting in inappropriate reliance, either rejecting correct suggestions or accepting suboptimal ones~\cite{Lfstrm2023OnTD}.
This risk is especially pronounced with 2D minimaps, as these interfaces require users to mentally map the information to their corresponding physical environment, 
imposing \carton{substantial} cognitive load due to reference-frame switching~\cite{Harris2013UnderstandingST}.
In contrast, embedded visualizations can \carton{substantially} reduce or eliminate these spatial mapping demands by directly fusing information within the physical environment,
thus potentially fostering more \emph{appropriate reliance} on AI suggestions.

Based on this rationale, we hypothesize that in AI-aided spatial decision-making, compared to the 2D minimap:
\begin{itemize}[leftmargin=1cm] 
     \item[\textbf{H1}] AR X-ray improves overall task performance, specifically decision accuracy.
    \item[\textbf{H2}] AR X-ray promotes more appropriate reliance on AI. 
    \item[\textbf{H3}] AR X-ray shortens response times in decision-making.
\end{itemize}
\section{Results}

\subsection{Quantitative Results}
Prior to analysis, we excluded trials in which participants failed to make a decision within 20 seconds. 
After filtering, we retained 248 trials in \mapAI, 247 in \mapnoAI, 237 in \ARAI, and 237 in \ARnoAI, out of 1024 (32 participants $\times$ 32 trials). \carton{For each measure, trials were aggregated by participant and conditions being analyzed.} We first tested for normality using the Shapiro-Wilk test. 
Results indicated that accuracy, point error, and AI reliance significantly violated the assumption of normality, whereas time and confidence did not. 
Accordingly, we used repeated-measures ANOVA ($\alpha = .05$) to assess the significance effects on time and confidence; then, post-hoc pairwise comparisons were conducted using Holm-Bonferroni corrections. 
For accuracy and point error, we used Aligned Rank Transform (ART) ANOVA~\cite{Wobbrock2011ART} to test significance and use ART-C~\cite{Elkin2021ART_C} for post-hoc contrast testing. 
To compare AI reliance in the \mapAI and \ARAI conditions, we used Wilcoxon signed-rank tests.

\begin{figure}[h]
    \centering
    \includegraphics[width=\linewidth, keepaspectratio]{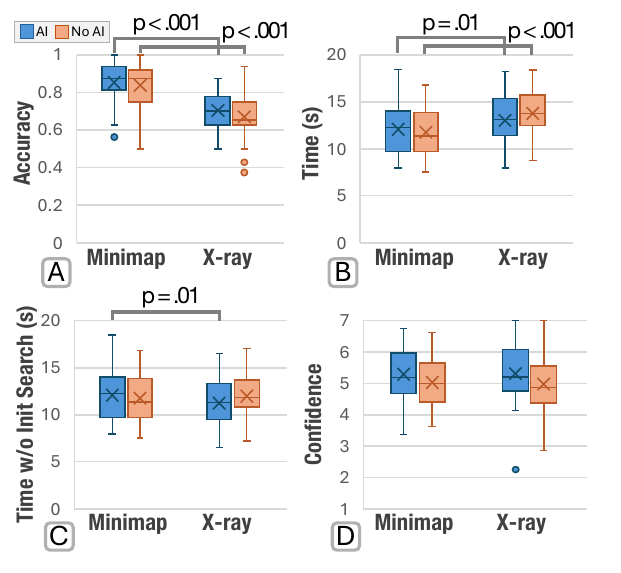}
    \vspace{-4mm}
    \caption{Summary of quantitative results across conditions. Metrics include (A) decision accuracy, (B) task completion time, (C) completion time excluding initial search, and (D) confidence ratings. Comparisons are shown between AI-assisted and non-AI conditions for both \map and \AR visualizations, with significance levels indicated. \carton{The ``X'' denotes the mean, the horizontal line inside each box indicates the median, the box spans the interquartile range (IQR). Outliers are defined as values beyond the upper cutoff $Q_3 + 1.5 \times \mathrm{IQR}$ or the lower cutoff $Q_1 - 1.5 \times \mathrm{IQR}$.}}
    \Description{Summary of quantitative results across conditions. Metrics include (A) decision accuracy, (B) task completion time, (C) completion time excluding initial search, and (D) confidence ratings. Comparisons are shown between AI-assisted and non-AI conditions for both \map and \AR visualizations, with significance levels indicated. \carton{The ``X'' denotes the mean, the horizontal line inside each box indicates the median, the box spans the interquartile range (IQR). Outliers are defined as values beyond the upper cutoff $Q_3 + 1.5 \times \mathrm{IQR}$ or the lower cutoff $Q_1 - 1.5 \times \mathrm{IQR}$.}}
    \label{fig:QResult}
\end{figure}

\subsubsection{Decision Accuracy}
\label{sec:acc}
We computed the mean accuracy of target selection for each condition \carton{at the participant level by averaging across all valid trials for each participant} (\autoref{fig:QResult}a).
The analysis showed a significant effect between \textit{Visualizations} on the decision accuracy ($F(1, 93) = 82.07, p < .001$). Post-hoc analysis further revealed that participants were more accurate in the \mapAI condition ($Mdn = 0.88, IQR = 0.13$) than in \ARAI ($Mdn = 0.70, IQR = 0.13$), with a significant effect ($F(1,93) = 6.26, p < .001$). Participants were more accurate in \mapnoAI ($Mdn = 0.88, IQR = 0.16$) than in \ARnoAI ($Mdn = 0.66, IQR = 0.13$), with a significant effect ($F(1,93) = 6.56, p < .001$). These results \textbf{reject our first hypothesis H1}.

Further observation suggests that \textbf{\ARAI under-performance may stem from inappropriate reliance on AI}.
Participants in the \ARnoAI condition achieved a median decision accuracy of $Mdn = 0.66$, which increased to $Mdn = 0.70$ in the \ARAI condition. However, this improvement still \emph{fell short of} the AI standalone baseline accuracy of 0.75, suggesting that participants may have relied on it inappropriately.
This aligns with findings in prior empirical studies~\cite{2021BansalDoesTheWholeExceedItsParts}.
In contrast, performance in both \mapnoAI ($Mdn = 0.88$) and \mapAI ($Mdn = 0.88$) conditions exceeded the AI baseline, indicating that participants were able to appropriately rely on AI by identifying and overriding poor suggestions.

% based on whether the AI's suggestion was correct or incorrect.
To further examine the nature of participants' reliance, specifically whether it reflected under- or over-reliance, we broke down trials by AI correctness into AI-correct, no-AI, and AI-wrong. The analysis indicated there is a significant effect of \textit{AI Correctness} ($F(2,155) = 47.97, p < .001$), \textit{Visualizations} ($F(1, 155) = 73.36, p < .001$), and \textit{AI Correctness}$\times$\textit{Visualizations} interaction ($F(2, 155) = 3.43, p =0.03$) on decision accuracy. Post-hoc analysis further revealed that there are significant differences in the \ARAI condition. Compared to the \ARnoAI baseline, participants  treated AI input as authoritative: they performed better when the AI was correct ($\Delta_{Mdn} = +0.20, F(1, 155) = 4.76, p < .001$), but worse when the AI was wrong ($\Delta_{Mdn} = -0.16, F(1, 155) = -3.37, p = .01$). 
This result indicated that participants tended to defer their decisions to the AI, even when the AI is misleading.
In contrast, for \map, we found no significant differences between AI-correct and no-AI conditions, nor between AI-wrong and no-AI conditions.
In sum, the breakdown analysis indicated that \textbf{participants exhibited greater inappropriate reliance on AI in \ARAI compared to \mapAI}.
Next, we further analyzed reliance patterns using dedicated reliance metrics.

\begin{figure}[htb]
    \centering
    \includegraphics[width=\linewidth, keepaspectratio]{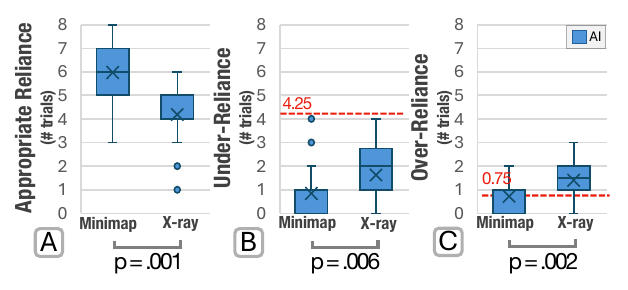}
    \vspace{-4mm}
    \caption{Number of trials per block in which participants exhibited each type of AI reliance, including (A) appropriate reliance, (B) under-reliance, and (C) over-reliance. 
    Random-selection baselines (in red) are indicated. \carton{The ``X'' denotes the mean, the horizontal line inside each box indicates the median, the box spans the interquartile range (IQR). Outliers are defined as values beyond the upper cutoff $Q_3 + 1.5 \times \mathrm{IQR}$ or the lower cutoff $Q_1 - 1.5 \times \mathrm{IQR}$.}}
    \Description{Number of trials in which participants demonstrated different types of AI reliance. The types include (A) appropriate reliance, (B) under-reliance, and (C) over-reliance, comparing \mapAI and \ARAI conditions. 
    Random-selection baselines (in red) are indicated. \carton{The ``X'' denotes the mean, the horizontal line inside each box indicates the median, the box spans the interquartile range (IQR). Outliers are defined as values beyond the upper cutoff $Q_3 + 1.5 \times \mathrm{IQR}$ or the lower cutoff $Q_1 - 1.5 \times \mathrm{IQR}$.}}
    \label{fig:Q_Rely_Result}
\end{figure}

\subsubsection{AI Reliance}
Results from the accuracy analysis indicated that participants relied on AI inappropriately in the \ARAI condition.
Quantitative analysis on the AI Reliance measure confirms this pattern:
we observed a significant difference between \mapAI and \ARAI, 
with participants showing less appropriate reliance on AI in \ARAI (\autoref{fig:Q_Rely_Result}a).
This finding again \textbf{contradicts our second hypothesis H2}.

Inappropriate reliance on AI can be either over-reliance or under-reliance.
To examine which pattern dominated participants' interactions with AI in \ARAI,
we separately analyzed trials of over-reliance and under-reliance. 
\autoref{fig:Q_Rely_Result}b and c depict the average occurrences of over- and under-reliance, respectively, per block (8 trials). 

To interpret these numbers, we compared the observed behaviors with a random-selection baseline. 
Among the 8 trials in each block, 
the AI suggested 5 optimal, 2 sub-optimal, and 1 worst target.
If a participant were to choose a target at random, \final{namely} with equal probability of $1/4$ for each of the four possible targets, the expected numbers of over- and under-reliance trails per block are 0.75 and 4.25 respectively (see Supplemental Materials for calculation details).
Compared to this baseline,
participants in the \ARAI condition displayed fewer instances of under-reliance 
($M_{\text{\AR}} = 1.63 < M_{\text{random}} = 4.25$), 
but notably more instances of over-reliance ($M_{\text{\AR}} = 1.41 > M_{\text{random}} = 0.75$).
In contrast, participants in 
the \mapAI condition exhibited fewer instances of both under-reliance ($M_{\text{\map}} = 0.84 < M_{\text{random}} = 4.25$) and over-reliance ($M_{\text{\map}} = 0.72 < M_{\text{random}} = 0.75$).
These findings indicate that \textbf{inappropriate reliance in the \ARAI condition was primarily driven by over-reliance}, consistent with our earlier accuracy analysis (Sec.~\ref{sec:acc}).

\subsubsection{Response Time}
% We computed the average response time across all trials in different conditions 
We computed the average response time across conditions \carton{at the participant level}, and observed significant effects (\autoref{fig:QResult}b). The analysis showed a significant effect between \textit{Visualizations} ($F(1,31) = 44.30, p < \text{.001} $) and \textit{Visualizations} $\times$ \textit{AI Conditions} interaction ($F(1, 31) = 5.92, p = \text{.02} $) on the response time. Further post-hoc analysis showed that participants responded faster in \mapAI ($M = 12.08, SE = 0.48$) than \ARAI ($M = 13.03, SE = 0.48$) with a significant effect ($\Delta_{M} = \text{-0.94}, F(1,31) = -3.12, p = \text{.01}$), and also in \mapnoAI ($M = 11.76, SE = 0.44$) than \ARnoAI ($M = 13.77, SE = 0.44$) with a significant effect ($\Delta_{M} = \text{-2.00}, F(1,31) = -6.28, p < \text{.001}$).
These results \textbf{contradict our third hypothesis H3}.

\para{Response Time Excluding Initial Search.}
We observed that 
participants in the \AR conditions spent additional time to visually search the spatially distributed targets—especially those outside their immediate field of view.
In contrast, in the \map{} conditions, 
all targets were presented in a 2D overview, making search overhead negligible.
Prior decision-making model (Simon's model~\cite{Simon1977}) and empirical studies ~\cite{reisen2008identifying,russo1994eye} suggest that actual \textit{decision-making} begins only after sufficient information has been gathered. Thus, similar to prior research ~\cite{Jakobsen2013VISonDisplay}, we distinguished the \emph{initial search} phase from the subsequent \textit{decision-making} phase.
This motivates us to exclude the initial search overhead in the \AR conditions
to better compare its \textit{decision-making} phase with the \map conditions.

\begin{figure}[htb]
    \centering
    \includegraphics[width=\linewidth, keepaspectratio]{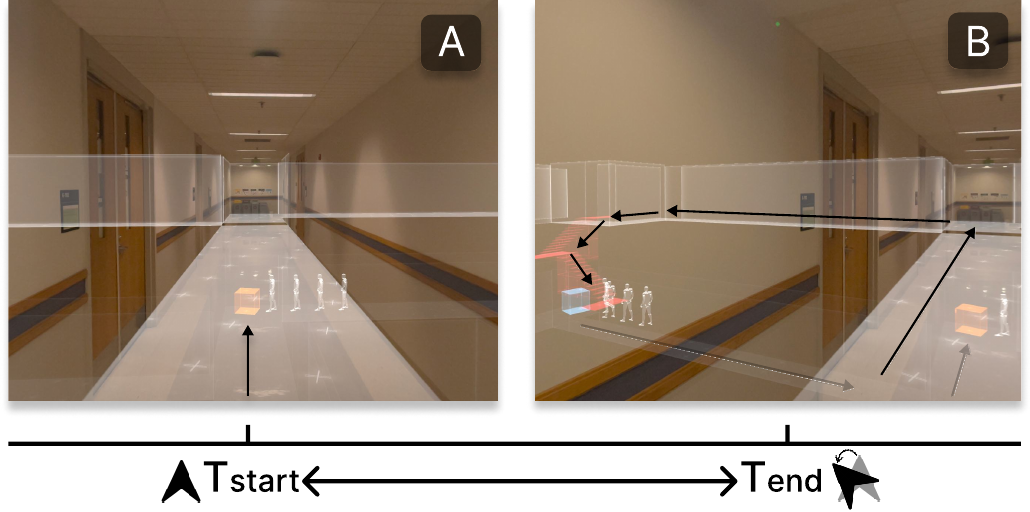}
    \vspace{-4mm}
    \caption{(a) and (b) show the direct and walking distance to the orange target; (a) and (b) also indicate the start and end moment of searching for the blue target.}
    \label{fig:DirectAndWalking}
    \Description{(a) and (b) show the direct and walking distance to the orange target; (a) and (b) also indicate the start and end moment of searching for the blue target.}
\end{figure}

Following prior visualization-decision work~\cite{Padilla2018DecisionMW}, 
we consider that meaningful decision-making start only after at least two options have been identified.
Thus, for each trial in the \AR{} conditions, we excluded the
initial search time required to discover \textbf{\emph{the first two distinct targets}}.
Specifically, we defined two search segments to exclude:
S1-from the start of the trial to the first frame when the first target entered the user's field of view; and 
S2-from the frame when the user turned away from the first target (\autoref{fig:DirectAndWalking}a) to the frame when the second target appeared (\autoref{fig:DirectAndWalking}b).
If a target was already visible at trial start, or if the first and second targets appeared simultaneously, the corresponding segment (S1 or S2) was set to zero.
This definition provides a strict lower bound on search time, 
as participants may still require additional head movement to center the target in view. 
Yet,
we acknowledge that this measure is not a theoretically perfect indicator of true decision time. Nevertheless, it provides a scientifically reasonable approximation that is consistent with prior research \final{~\cite{Jakobsen2013VISonDisplay}}. We discuss its limitations and possible refinements in Sec.~\ref{sec:limitation}.

Two co-authors independently annotated these segments frame by frame from the first-person recordings to ensure inter-rater reliability. \carton{Inter-rater reliability was assessed using a two-way mixed effects intraclass correlation coefficient, ICC(3,1), which indicated excellent agreement ($\mathrm{ICC}(3,1) = 0.995$, 95\% CI [0.992, 0.997])~\cite{koo2016guideline}. Encoding consistency was ensured through frame-level accurate marking and the clear visual cues afforded by first-person video recordings.} Discrepancies were discussed and resolved through consensus. After subtracting the annotated initial search time from the response time, we observed a different pattern (\autoref{fig:QResult}c). The analysis showed that there is a significant effect of \textit{Visualizations} $\times$ \textit{AI Conditions} interaction ($F(1,31) = 7.70, p = \text{.009}$). Further post-hoc analysis indicated that participants in \ARAI 
($M = 11.23, SE = 0.48$) 
spent less time than those in \mapAI 
($M = 12.08, SE = 0.48$) 
with a significant effect ($\Delta_{M} = \text{-0.85}, F(1,31) = -3.27, p = \text{.016}$),
indicating that, \textbf{with AI assistance, decisions were made more quickly with the X-ray visualization}.
No significant difference was observed between \mapnoAI and \ARnoAI,
suggesting that, without AI assistance, participants spent comparable time making decisions across visualizations.
In summary, 
although raw response times were longer in \AR conditions,
a detailed breakdown suggests that this extra time wasn't entirely spent on the decision-making.
This leads to interesting implications that we will further discuss in Sec.~\ref{sec:reduce}.

\subsubsection{Post-Trial Pointing Error Rate}
We analyzed participants' post-trial pointing error rates across the two visualizations (\map vs. \AR) \carton{at the participant level}, regardless of AI availability.
We found a significant effect ($ F(1, 93) = 40.77, p < .001$), with higher error rates in the \map condition ($Mdn = 0.08, IQR = 0.20$) compared to the \AR condition ($Mdn = 0, IQR = 0$).
This result indicates that \textbf{the X-ray visualization significantly improved participants' spatial mapping ability compared to the Minimap}.

\subsubsection{Self-Reported Confidence}
We also examined participants' self-reported confidence ratings across the four conditions.
This measure reflects how confident participants were that they had selected the optimal target in each trial.
Our analysis revealed no significant differences across conditions (\autoref{fig:QResult}d),
suggesting that \textbf{participants felt equally confident in their decisions regardless of visualization type or AI assistance}.

\subsection{Qualitative Results}
We conducted semi-structured post-study interviews to capture participants' strategies, rationales, and motivations behind their decision-making across all conditions. Participants are referred to as P1–P32. 
We analyzed the interviews using reflexive thematic analysis~\cite{braun2019reflecting}, 
identifying themes both inductively (bottom-up) and deductively (top-down).
Below, we summarize the key themes that emerged from the analysis.

\subsubsection{Diverse Decision Strategies Across Visualizations}
We observed that participants adopted different decision-making strategies depending on the visualization.

\para{Both Visualizations: People First and Distance First Strategies.}
Participants primarily adopted one of two strategies:
(1) prioritizing fewer people in line, then considering walking distance; or
(2) prioritizing shorter distance, then checking queue length.
In the \map conditions, participants were almost evenly split between these two strategies, 
with roughly half employing a ``people first'' (15/32) and the other half a ``distance first'' (14/32).
In the \AR condition, more participants adopted a distance-first strategy (15 out of 32), while fewer prioritized queue lengths (11/32). 
The remaining participants (6/32) reported balancing both factors equally without a dominant criterion.

\para{X-Ray: Unique Heuristics and Motivations.}
% --- VIS25 version ---
% Whereas strategy selection in the \map conditions appeared to depend largely on participants' spatial  skills, the \AR conditions introduced distinct perceptual heuristics influencing participants to prioritize distance:
While strategy in the \map condition appeared to depend largely on participants' spatial skills, the \AR condition introduced distinct perceptual heuristics influencing participants to prioritize distance:
\begin{itemize}[noitemsep]
    \item \textbf{Egocentric Spatial Imagery.} 
    The egocentric view in our AR \carton{X-ray} appeared to support participants in vividly imagining physical navigation, particularly for complex pathways involving hallway turns or floor changes. For example, P9 noted, \quot{It's easier to imagine how I should walk to the coffee machine, especially when targets are on a different floor from me...}
    Such imagination of navigation in the AR \carton{X-ray} may have led to weigh distance more heavily in their decisions.
    % This detailed mental imagination caused them to weigh distance more heavily in their decisions.
    
    % P5 noted, \quot{ ... you are trying to get there [in AR interface]}
    % . Imagine a path there or even just a straight line in AR whereas in minimap it gets difficult because they're two floors apart sometimes. AR seems more realistic.}

        % --- VIS25 version ---

    % 
    % This illusion occasionally influenced participants to overweight distance.

    \item \textbf{Visual Proximity Illusions.} 
Participants reported a misleading proximity effect caused by the X-ray view’s “see-through” nature.
Because the visualization allows users to see targets behind walls, some targets appeared directly nearby (\autoref{fig:DirectAndWalking}a) even though the actual walking path to reach them was much longer (\autoref{fig:DirectAndWalking}b).
This discrepancy often led users to underestimate the true walking distance, particularly when they cannot easily find the path. 
As P8 noted, \quot{I think this way is shorter but it's actually longer.}
    This difficulty occasionally placed additional cognitive burden on participants and may led them to overweight distance.

    % As P12 noted, \quot{I was just looking for the closest one. But I am not sure if I can get to it because there is no way. The closest one only looks close from direct distance, but the walking distance may not be close.}.    
    % [P12] noted, \quot{For AR, actually, I was just looking for the closest one. But I'm not sure because probably I can't get to it because there is no way. The closest one only looks close from direct distance, but the walking distance may not be close."}
    % [P23] noted, \quot{For that I feel like I just picked whichever was more direct and I was sure that I could get there fast.}
    
    \item \textbf{Occlusion Challenges.} 
    Some participants \final{such as} P2, P7 and P11 reported difficulties estimating queue lengths due to occlusion in the \AR conditions, prompting reliance on distance estimates instead of people counts.
    % noted, \quot{I couldn't clearly see how many people were waiting because avatars overlapped, so I just relied on how close it seemed}). 
    % See Figure ~\autoref{fig:PplOverlap}.
\end{itemize}

\begin{figure}[htb]
    \centering
    \includegraphics[width=0.8\linewidth, keepaspectratio]{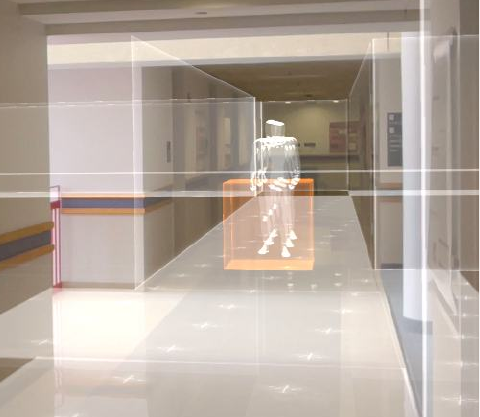}
    \caption{Example of occlusion in the AR \carton{X-ray} visualization, where overlapping avatars make it difficult to estimate the number of people in the queue, leading to potential misjudgment of wait time.}
    \Description{Example of occlusion in the AR \carton{X-ray} visualization, where overlapping avatars make it difficult to estimate the number of people in the queue, leading to potential misjudgment of wait time.}
    \label{fig:PplOverlap}
\end{figure}

\para{Summary.}
Compared to the \map conditions, participants in the \AR conditions showed a stronger preference for spatial reasoning (distance) in their decision-making. This suggests that the AR \carton{X-ray} visualization inherently facilitates spatial perception, yet they also function as a double-edged sword by introducing perceptual biases and occlusion challenges.
% --- VIS25 version ---
% This suggests that the AR embedded visualizations inherently facilitate spatial perception, which, however, can be a double-edged sword that introduces perceptual biases and occlusion challenges.

\subsubsection{Increased Trust in AI Suggestions with X-Ray}
The visualizations also impacted participants' subjective trust on AI.

\para{\map: Make Own Choice First, Then Consult AI.}
In the \map conditions, most participants (22 out of 32) preferred first to form their own judgment 
and then cross-check against the AI suggestion, \carton{for instance,} \quot{I use it just to double-check my decision} (P22). 
Participants reported feeling more confident in their own judgment when using the \map visualization because it presented the information required to make informed choices. 
They typically consulted the AI in cases of uncertainty or indecision:
% Participants turned to AI primarily in cases of uncertainty, indecision, or complexity between options 
\quot{If I felt stuck between two locations, I would check the AI to break the tie} (P20).

\para{\AR: AI as a Starting Anchor.} 
Participants expressed greater trust in AI suggestions in the \AR conditions, 
frequently using the AI suggestion as an initial reference or anchor for their decisions.
This subjective feedback may help explain the observed over-reliance.
Several participants \carton{such as} P6 and P11 mentioned that AI helped narrow the decision space and enabled quicker comparisons.
% Participants described this as narrowing the decision space and 
% allowing quicker and easier comparisons (\eg, P6, P11).
% (\quot{AI narrowed down my choices quickly; I just needed to verify if anything was obviously better} [P6 and P11]).
Participants provided several reasons for trusting AI more with \AR:
\begin{itemize}[noitemsep]
    \item \textbf{Perceptual and Navigation Challenges.} 
    Participants found distance estimation (P2),
    % (\quot{It's hard to gauge distances accurately} [P2]), 
    queue length assessment (P4, P5),
    % (\quot{Can’t easily tell how many people are waiting} [P2,P4,P5,and P11]), 
    and locating navigation paths (P12) more effortful in AR.
    % \quot{I can't find the path fast enough in AR}[P12] \quot{I couldn't really see how many hallways are in between.} [P3]). 
    Thus, AI suggestions offered a useful cognitive shortcut, 
    aligning with findings from prior research on AI-aided decision-making~\cite{2021BansalDoesTheWholeExceedItsParts}.

    \item \textbf{Unfamiliarity with AR X-ray.}
    Some participants \carton{like P31} mentioned that \carton{unfamiliarity} with the AR X-ray increased their reliance on AI guidance.
    % (\eg, P3 \quot{I'm not yet comfortable with how AR displays information...AI really helps there} [P31]).

    \item \textbf{Enhanced Immersive Realism and Embodiment of AI suggestions.}
    Interestingly, several participants \carton{such as P4 and P13,} indicated that the realistic, immersive, embodied presentation of AI suggestions within the AR X-ray heightened their sense of trust and acceptance, which sometimes led to over-reliance on wrong AI suggestions.
    % (\quot{The way AR showed the AI’s pick felt intuitive and believable} [P13, P20 and P21]).
\end{itemize}

\para{Summary.} 
Compared to the \map, the \AR visualization led to greater participant trust in and reliance on AI suggestions. 
Participants attributed this shift to greater perceptual complexity, unfamiliarity with the AR \carton{X-ray} visualization, and the more realistic presentation of AI.

\section{Discussion}

In this section, we synthesize the key 
lessons learned,
design implications,
and promising future directions
to advance human-AI collaboration in spatial tasks in AR.

\subsection{Perceptual Challenges of AR X-Ray Can Lead to Over-Reliance on AI}
Our findings suggest that perceptual challenges of the AR \carton{X-ray}, such as occlusion, large canvas space, and spatial estimation difficulty, can increase users' reliance on AI suggestions, 
even when those suggestions are suboptimal. 
This over-reliance does not appear to stem from blind trust in AI, 
but rather from the inherent complexity of perceiving and evaluating visual information in large-scale physical environments.
While prior research has discussed AR's perceptual limitations~\cite{Drascic1996PerceptualII}, 
our study directly connects these challenges to AI reliance patterns in time-sensitive spatial decision-making. This highlights the need for better embedded visualization designs that support spatial reasoning, especially in large-scale environments.

\para{Design Implications.}
Visually embedding AI support in physical environments is not inherently beneficial. 
It can be neutral or even detrimental when perceptual challenges make it harder to verify AI suggestions. 
These findings underscore the need to rethink how AI assistance is integrated into AR, 
rather than directly porting AI designs from desktop or 2D interfaces. 
Designers should explicitly consider how spatial perception difficulties affect users' ability to evaluate AI output.

\para{Future Studies.}
In our study, situational data \carton{like} queue length was visualized using human-shaped glyphs.
A natural possibility is to compare visual encoding (e.g., glyphs vs. numeric labels) to evaluate how data representation \final{like} legibility, uncertainty and saliency influences AI reliance. 

Additionally, examining how different fundamental characterizations of the situation, \final{such as} task difficulty, information ambiguity and time pressure, modulate trust and reliance in AR embedded visualizations, would offer deeper insight into AR-AI interaction design.

\subsection{Cognitive Biases Triggered by AR X-Ray Can Induce Inappropriate AI Reliance}
Beyond perceptual difficulty, our qualitative findings point to cognitive biases that are unique to the AR \carton{X-ray}. 
Participants reported visual proximity illusions and increased trust in AI suggestions due to their spatial embodiment in the environment. These effects align with prior research on spatial cognition biases~\cite{raghubir1996crow} and trust in embodied agents~\cite{Kim2018DigitalAssistantNeedABody}.

Although cognitive biases have been studied in traditional visualization contexts~\cite{dimara2018task, wall2021left}, 
their implications in embodied, spatial, and AI-augmented AR systems remain underexplored. 
As Padilla \etal~\cite{padilla2018decision} suggested, when systems are designed with such biases in mind, 
users can benefit from Type 1 (intuitive) thinking—particularly under time pressure. 
Our findings highlight a fertile area for interdisciplinary research at the intersection of spatial cognition, embedded visualizations, and human-AI decision-making.

\para{Design Implications.}
Embedded visualizations in AR should explicitly account for and possibly counteract spatial cognitive biases. 
For example, visual proximity illusions may be mitigated through alternative depth cues, or warnings when direct-line views differ from walking paths. 
Designers might also consider using less embodied, realistic representation for AI suggestions 
when the risk of over-reliance is high.

\para{Future Studies.}
While spatial AR studies are often limited by physical constraints of real-world spaces, 
 the increasing accessibility of AR and AI-integrated devices makes spatial cognition experiments increasingly feasible. 
Future research should investigate how AR-induced biases interact with varying levels of AI accuracy or explanation quality, particularly in high-stakes or time-sensitive scenarios. 
This represents a promising interdisciplinary frontier spanning visual computing, cognitive science, and human-computer interaction.

\subsection{Aligning the Strengths of AR X-Ray with Spatial Tasks}
In our study, while the X-ray did not outperform the Minimap on decision accuracy, 
it significantly reduced post-trial pointing errors, 
suggesting improved spatial mapping. 
This distinction suggest that AR \final{X-Ray} \carton{may} be more beneficial in continuous, action-based spatial workflows rather than in isolated decision-making tasks.

In spatial tasks like emergency response, selecting a correct location is only the first step; what follows is action. 
Misinterpreting a target's location after selection can be consequential. 
For instance, a user might correctly choose an exit from a wall-mounted map but walk the wrong way due to reference-frame confusion. 
AR \final{X-Ray} mitigates this risk by spatially anchoring information in the environment.
Thus, rather than emphasizing the weakness of AR \final{X-Ray}, our findings illustrate their potential strengths in embodied spatial tasks and highlight when and where AR embedded visualizations are best applied.

\para{Design Implications.}
AR embedded visualizations \carton{may} be most effective when tightly coupled with tasks involving physical action, rather than static decision-making.
Designers should consider the viewer's movement and real-time orientation when designing the embedded visualizations.
Emerging work on ``Visualization in Motion''~\cite{DBLP:journals/IEEE_TVCG/YaoBVI22} and ``First-Person View
Visualization''~\cite{isenberg:hal-04782928} provides valuable guidance in this direction.
In addition, AR systems should support smooth transitions between embedded visualizations and traditional 2D visualizations in the headset depending on task context and user needs.

\para{Future Studies.}
In our study, participants selected targets verbally.
A natural follow-up would involve embodied selection, \carton{for example}, walking to or pointing at targets, to examine whether the spatial support of AR embedded visualizations becomes more critical under physical interaction.
We hypothesize that in such embodied tasks, participants using 2D maps will incur greater cognitive load in spatial mapping, possibly diminishing their ability to evaluate targets effectively. 
Future work should explore how task modality, \final{such as} static vs. embodied, interacts with visualization and AI support in complex environments.

\subsection{AR X-Ray May Lower Cognitive Engagement Compared with 2D Visualizations}
\label{sec:reduce}
In our study, there was no significant difference in response time between the \mapnoAI and \ARnoAI conditions once initial search time was excluded. 
However, participants in the \ARAI condition made decisions more quickly yet achieved lower overall accuracy than those in the \mapAI condition. 
This pattern aligns with reliance on heuristic judgments, often described as Type 1 thinking~\cite{Chaiklin2012ThinkingFA, JSBTDualProcessTheory2013}. 
Together with the high self-reported confidence, these results suggest a lack of deeper cognitive engagement: once AI assistance was available, participants quickly felt confident enough in their intuitive judgments, rather than engaging in more analytical (Type 2) thinking.
These findings imply that AR \final{X-Ray}, by offering a highly immersive or embodied experience, \carton{may} inadvertently reduce users' motivation to scrutinize information.

\para{Design Implications.}
 It remains unclear whether this lower cognitive engagement is specific to our task design, a broader characteristic of AR embedded visualizations, or a more general cognitive shift from 2D to AR egocentric perspective. Design strategies that prompt more deliberation, such as ``cognitive forcing functions''~\cite{2021ToTrustOrToThink}, could be incorporated to encourage deeper analytic thinking in AR environments.

\para{Future Studies.}
Further research is needed to determine whether AR embedded visualizations consistently reduce cognitive motivation compared to 2D visualizations, and to explore methods for mitigating AR-induced cognitive shortcuts across diverse tasks, visualization designs, and populations.

\subsection{Study Limitations}
\label{sec:limitation}
Our study offers an initial examination of AI-aided spatial decision-making with AR embedded visualizations. Although guided by prior research on AR visual systems for spatial tasks~\cite{XUImroveIndoorWayfinding, XUARteambasedsearch}, 
we acknowledge that our system does not represent a fully optimized, advanced, or universal solution. 

To ensure experimental control, we adopted fixed parameters, such as a discrete set of spatial arrangements, an AI accuracy of 75\% (see Section~\ref{sec:simulated_AI}), and a 20-second decision window, which may not capture the full complexity of high-stakes, real-world environments or more advanced AI systems. 
Consequently, our findings are most transferable to tasks resembling the indoor, time-sensitive context described in Section~\ref{sec:task}.
Moreover, our participant pool primarily comprised well-educated individuals, which may not reflect the behavior of specialized user groups, \carton{for example}, emergency responders or security professionals. 
As is common in spatial cognition research~\cite{Tan2024InvisibleME}, our sample size was modest; although sufficient for controlled experimentation, a larger and more diverse sample would strengthen external validity. 

Finally, in our analysis, there may be a minor initial search time in the \map condition, which depends on how quickly participants detect the first two targets in the Minimap. 
However, this subtle interval cannot be reliably measured from first-person video alone. Future work incorporating eye-tracking data, which is currently unavailable on Apple Vision Pro, could help capture this duration more accurately.
We also acknowledge that during S2, some cognitive evaluation of the first target may occur. Nevertheless, by definition, a decision can only be made once the second target has been located.
Overall, we view this study as a scientifically grounded first step that provides a reasonable and well-documented foundation for subsequent research to extend and refine AR-based decision support across broader tasks and populations.

\section{Conclusion}
In this paper, we investigated the impact of AR embedded visualizations on human-AI reliance in spatial decision-making under time constraints by comparing X-ray with Minimap. 
Our user study (N = 32) revealed that the \carton{X-ray} visualization tends to evoke over-reliance on AI during spatial decision-making. 
This effect may be linked to perceptual challenges \final{like} occlusion and distance estimation, and AR-specific biases such as visual proximity illusions and the heightened sense of realism and immersion induced by high-fidelity visuals.
Nonetheless, the \carton{X-ray} visualization exhibits unique strengths by enhancing spatial \carton{mapping}, as evidenced by significantly reduced pointing errors.
We hope that our findings and design implications will guide future research and the development of solutions to better leverage AR's advantages to improve human-AI collaboration in spatial contexts.

%%
%% The acknowledgments section is defined using the "acks" environment
%% (and NOT an unnumbered section). This ensures the proper
%% identification of the section in the article metadata, and the
%% consistent spelling of the heading.
\begin{acks}
The author(s) would like to express sincere gratitude to the anonymous reviewers for their constructive suggestions.
\end{acks}

%%
%% The next two lines define the bibliography style to be used, and
%% the bibliography file.
\bibliographystyle{ACM-Reference-Format}
\bibliography{reference}

%%
%% If your work has an appendix, this is the place to put it.
% \appendix

% \section{Research Methods}

% \subsection{Part One}

% Lorem ipsum dolor sit amet, consectetur adipiscing elit. Morbi
% malesuada, quam in pulvinar varius, metus nunc fermentum urna, id
% sollicitudin purus odio sit amet enim. Aliquam ullamcorper eu ipsum
% vel mollis. Curabitur quis dictum nisl. Phasellus vel semper risus, et
% lacinia dolor. Integer ultricies commodo sem nec semper.

% \subsection{Part Two}

% Etiam commodo feugiat nisl pulvinar pellentesque. Etiam auctor sodales
% ligula, non varius nibh pulvinar semper. Suspendisse nec lectus non
% ipsum convallis congue hendrerit vitae sapien. Donec at laoreet
% eros. Vivamus non purus placerat, scelerisque diam eu, cursus
% ante. Etiam aliquam tortor auctor efficitur mattis.

% \section{Online Resources}

% Nam id fermentum dui. Suspendisse sagittis tortor a nulla mollis, in
% pulvinar ex pretium. Sed interdum orci quis metus euismod, et sagittis
% enim maximus. Vestibulum gravida massa ut felis suscipit
% congue. Quisque mattis elit a risus ultrices commodo venenatis eget
% dui. Etiam sagittis eleifend elementum.

% Nam interdum magna at lectus dignissim, ac dignissim lorem
% rhoncus. Maecenas eu arcu ac neque placerat aliquam. Nunc pulvinar
% massa et mattis lacinia.

\end{document}